\documentclass{aa} 
\usepackage{graphicx,subcaption} 
\usepackage{xcolor} 
\usepackage{amsmath} 
\usepackage{textgreek} 
\usepackage{caption, threeparttable} 
\usepackage{natbib} 
\usepackage{gensymb} 
\usepackage{tabularx} 
\usepackage{booktabs} 
\usepackage{array} 
\usepackage{float} 
\usepackage{txfonts} %
\usepackage{lipsum} 
\usepackage{multicol} 
\usepackage{mwe} 
\usepackage[font=small,labelfont=bf]{caption} 
\usepackage{float} 
\usepackage{placeins} 
\usepackage[colorlinks=true,urlcolor=blue,citecolor=blue,linkcolor=blue,bookmarks=true]{hyperref}

\titlerunning{}
\authorrunning{Renu et al.}

\begin{document}

\title{Tracing evolutionary pathways of bar-driven quenching in local Universe disc galaxies}

   \author{D. Renu
          \inst{1,2}, 
          Smitha Subramanian\inst{1,2,3},
         \and
          Koshy George\inst{4}
          }
   \institute{Indian Institute of Astrophysics, 2nd Block Koramangala, Bengaluru, 560034\\
             \email{renukadyan3232@gmail.com, renu.devi@iiap.res.in} 
         \and
            Pondicherry University, R.V. Nagar, Kalapet, 605014, Puducherry, India
        \and
        Leibniz-Institut für Astrophysik Potsdam, An der Sternwarte 16, D-14482 Potsdam, Germany,
        \and
             Faculty of Physics, Ludwig-Maximilians-Universitat, Scheinerstr. 1, Munich 81679, Germany
             }

   \date{Received XXXX ; accepted XXXX}

\abstract
 {Bars play an integral role in regulating star formation in spiral galaxies, from triggering central starbursts to driving quenching. The diverse star formation morphologies observed in local barred galaxies reflect different evolutionary stages of the bar, making studies across these stages essential for understanding how bars regulate star formation. In this context, we study a sample of 12 nearby barred galaxies in the redshift range 0.01 – 0.06 to understand bar-driven quenching of star formation. These galaxies are identified as centrally quenched galaxies, those with extended star-forming discs but quenched inner regions, by leveraging the differences in star formation rates between the MPA–JHU and GSWLC catalogues. However, they also exhibit residual central emission within the region covered by the SDSS 3" fibre. The emission line analysis shows that the central emission in these galaxies originates from either ongoing star formation or LINER-like activity, suggesting diverse central ionization mechanisms. Based on the location of different structural components (bar, bulge, and disc) in spatially resolved UV–optical colour maps using SDSS (r-band) and GALEX (FUV and NUV) imaging data, we find that discs are star-forming and bluer in colour (NUV-r < 4 mag) while the bulge and bar regions are systematically redder (NUV-r > 4 mag) and dominated by older stellar populations. Following this, we derive NUV–r radial colour profiles (which is a good photometric age indicator) to constrain the ages of the dominant stellar populations across different regions of these galaxies. The profiles show a clear transition from red to blue colours at the bar end, with a corresponding median stellar age of $\sim$ 1 Gyr. 
We compare the NUV–r profiles of our sample with those of fully centrally quenched barred galaxies from our previous study, which show no emission within the 3" SDSS fibre. Although NUV-r remains > 4 mag inside the bar region, indicating stellar populations older than $\sim$ 1 Gyr, our galaxies are systematically bluer and younger at the same radii than the fully quenched sample.
This suggests that our current sample may represent an intermediate evolutionary stage in bar-driven quenching. To check the role of low-luminosity AGN activity in star formation quenching, we estimated the black hole masses for our sample. All galaxies lie below the threshold ($logM_{BH} < 8.0$) associated with kinetic-mode AGN feedback, implying that AGN-driven outflows are unlikely to be the primary quenching mechanism. Additionally, we also note that these galaxies, except one, host pseudo-bulges. These findings suggest that the most likely mechanism driving the observed quenching in our sample galaxies is the action of bars, and these sample galaxies represent an evolutionary phase in the bar-driven quenching process, just before the inner region is completely quenched. Higher spatial resolution observations of the bar and central sub-kpc region can help better constrain the bar-driven evolutionary pathways leading to the quenching of star formation in barred spiral galaxies.
}

   \maketitle

\section{Introduction}
\indent Bars have always been a fascinating component of galaxies that play a versatile role in galaxy evolution, from regulating star formation and gas inflows \citep{1993RPPh...56..173S, 2016A&A...595A..63V} to exchanging angular momentum with dark matter haloes \citep{2003MNRAS.341.1179A}. They are non-axissymmetric structures that redistribute gas from nearby regions toward the centre of galaxies, enhancing central gas densities which results in bursts of star formation  \citep{1995ApJ...454..623K, sheth2005secular, 2011ApJ...743L..13C, 2011MNRAS.416.2182E, 2021A&A...656A.133Q}, while also providing fuel for the central supermassive black hole \citep{1989Natur.338...45S, 2006LNP...693..143J, 2022A&A...661A.105S, 2024MNRAS.532.2320G}. Bar-driven gas flow from radii of several kpc down to the central few hundred pc, often results in the formation of nuclear structures, such as nuclear discs, nuclear bars, and nuclear rings (\citealt{2010MNRAS.402.2462C, 2014A&A...562A.121C, 2015ApJS..217...32B,2025MNRAS.544L..58G,2025A&ARv..33....7S} and references therein). Numerous simulations and observational studies have also shown that stellar bars can play a significant role in suppressing star formation \citep{gavazzi2015role, 2017A&A...598A.114C, khoperskov2018bar, james2018star, kruk2018galaxy, george2019insights, george2020more, 2019MNRAS.489.4992D, 2020MNRAS.492.4697N, 2024A&A...687A.255S}. This suppression of star formation, known as bar-quenching, has been reported to be particularly pronounced within the region covered by the length of the bar in the galaxy disc (bar region) \citep{james2009halpha,george2019insights, george2020more, george2021bar, 2025A&A...696A.118R}. Thus, bars can be one of the main drivers of internal quenching mechanisms in galaxies and could eventually lead to global quenching.  \\
\indent More than 70 per cent of disc galaxies in the local Universe are found to host stellar bars \citep{Eskridge_2000, nair2010fraction, masters2010galaxy, 2007ApJ...657..790M, 2018MNRAS.474.5372E, Lee_2019}. Studies using the Hubble Space Telescope (HST) and Atacama Large Millimeter Array (ALMA) have observed galaxies hosting bars at redshifts up to 1 -- 1.5 (\citealt{2003ApJ...592L..13S, 2008ApJ...675.1141S, 2014MNRAS.445.3466S, 2019ApJ...876..130H}). With the advent of deeper, high-resolution observations from missions such as the James Webb Space Telescope (JWST) and Euclid, it is now quite feasible to detect barred structures at even higher redshifts (\citealt{2024MNRAS.530.1984L,2025arXiv250315311E,2025ASPC..540...27P}. Recent studies using JWST reveal bar structures up to redshifts as high as z $\sim$ 4 (\citealt{2024arXiv240401918A, 2025ApJ...985..181G}). Though the bar fraction is found to decline to roughly 25–10 percent between redshifts z = 0.5 and z = 3 \citep{2025ApJ...985..181G, 2025ApJ...987...74G}, this fraction might represent a lower limit likely due to observational challenges in bar detection at higher redshifts. Recently, \cite{2025A&A...698A...5D} constrained the epoch of bar formation and estimated bar ages of up to $\sim$ 13 Gyr, suggesting that bar formation may occur very early in the Universe. \citet{2024A&A...684A.179R} using TNG50 simulation, have revealed that bars formed at high redshift (z $>$ 2) are dynamically long-lived and remain stable provided that strong environmental effects do not significantly perturb the host galaxy. These studies suggest that bars are found in galaxies across redshift, and can persist for several Gyr. Hence, they might be playing a crucial role in driving galaxy properties, including the quenching of star formation, at all redshifts. 

\indent Bars can suppress star formation by two physical mechanisms: either due to rapid gas consumption, depleting fuel for further star formation, or by making gas too turbulent to collapse and form stars. In the first mechanism, the bar exerts gravitational torques on the surrounding gas near its corotation radius, causing the gas to lose its angular momentum and flow inward toward the galaxy centre, triggering intense starburst activity. During this inflow phase, elevated levels of star formation may be seen along the bar, while simultaneously, the inner disc starts to get depleted of cold gas required for further star formation. Over time, the region between the bar’s corotation radius and the nucleus becomes gas-deprived and shows little to no ongoing star formation. Once the nuclear starbursts also consume the available gas, and if no further inflow is maintained, star formation in the central regions gradually diminishes, leaving the inner disc fully quenched. \cite{james2009halpha} \& \cite{2015MNRAS.450.3503J}, using H$\alpha$ radial profiles of barred galaxies, showed that star formation is strongly suppressed along the bar region, with enhanced activity only near the bar ends and centre. Similar distributions for gas and star formation have also been suggested in other studies like \cite{gavazzi2015role}, \cite{ spinoso2016bar}, \cite{george2019insights, george2020more}, and \cite{2024A&A...687A.255S}. The second mechanism, more apparent in gas-rich barred galaxies \citep{khoperskov2018bar}, is linked with bar-driven shocks and shear that can generate turbulence in the interstellar medium \citep{1998A&A...337..671R, Maeda_2023, 2022A&A...663A..61P, Kim_2024}. This induced turbulence increases the internal velocity dispersion of molecular clouds, inhibits their gravitational collapse, and prevents the formation of new stars. As a result, star formation is suppressed even in the presence of atomic and molecular gas, rendering the bar region in a quenched state. \cite{2015MNRAS.446.2468E} simulated a Milky Way-like barred galaxy to study bar formation and its evolution, and found that the star formation was concentrated only within nuclear regions inside the inner Lindblad resonance (a few hundred parsecs) and at the bar ends. However, along the bar itself, star formation was largely absent due to strong shear, which disrupted the gas clouds. These studies suggest that through either of the mechanisms described above or in combination, barred galaxies exhibit the suppression of star formation in the region between the nuclear sub-kpc region and the ends of the bar (region covered by the length of the bar in the galaxy disc).

\indent Whether or not star formation persists in the nuclear region likely depends on the different evolutionary stages of the bar evolution. Over time, as bars evolve and exhaust or redistribute available gas and stars, this suppression becomes more substantial, ultimately leading to a fully quenched central region, up to the extent of the bar. In addition to bar quenching, bar-driven gas inflows are also fuelling the central supermassive black hole, giving rise to LINER or AGN activity, which is observed more frequently in barred galaxies \citep{2024MNRAS.532.2320G, 2025A&A...699A.204M}. Such nuclear activity may also contribute to the suppression of nuclear or central star formation through heating or removal of gas. The formation of outer rings at co-rotation radii may further contribute to these processes by preventing the inflow of external gas into the central regions and replenishing fuel for further formation  \citep{2020ApJ...893...19W} and references therein). The centrally quenched barred galaxies may later transform into fully quenched galaxies if additional quenching mechanisms, such as environmental effects \citep{2012MNRAS.423.1485S} or the cessation of cosmological gas accretion \citep{2015A&A...580A.116G}, come into play. Thus, barred galaxies can display a wide range of star formation morphologies, depending on the evolutionary stage and timescale at which they are observed \citep{2020MNRAS.495.4158F, 2007A&A...474...43V, george2020more, 2020A&A...644A..38D}.

\indent Study of galaxies in different stages of bar evolution/quenching, can provide insights into the role of the bar in regulating the star formation in galaxies. In our previous work, D. \cite{2025A&A...696A.118R} (hereafter, RD25), we identified a sample of 17 centrally quenched barred galaxies in the redshift range (0.01 - 0.06) by leveraging the differences between two SFR estimates; from the  MPA-JHU (Max Planck Institute for Astrophysics and the Johns Hopkins University) Value Added Catalog (\cite{brinchmann2004physical}) and the GSWLC (GALEX-SDSS-WISE Legacy Catalog, \cite{2018ApJ...859...11S}). The centrally quenched galaxies are those galaxies with extended star-forming discs but with quenched inner regions. The criteria to select centrally quenched galaxies are explained in Section 2 of RD25. Using spatially resolved UV–optical colour maps, RD25 found that the bar and bulge regions in their sample galaxies were quenched, exhibiting NUV–r colours consistent with stellar populations older than 1 Gyr, whereas the outer discs showed recent star formation. These results supported a scenario in which bar-driven processes had ceased star formation within the bar region, including the sub-kpc nuclear region. Most of these galaxies host pseudo-bulges and do not host AGN, indicating that the most likely reason for the quenching of these galaxies (up to the extent of the bar) is the action of the stellar bar.

\indent In this current work, we study a sample of barred galaxies that are probable candidates of centrally quenched systems (up to the extent of the bar), but with emission features in the 3" SDSS fibre spectra. These systems offer a window into the stage where star formation may have ceased along the bar, but the presence of emission lines (like Halpha) in the SDSS spectra could be due to residual star formation in the central regions of the galaxy. We focus on these emission-line galaxies to test whether they represent an evolutionary phase in the bar-quenching sequence, where star formation is confined to the nuclear sub-kpc region, while the region between the bar ends and the center is already quenched. To determine whether the ionisation is due to star formation, we will first analyze the emission lines. Then, we will analyze their spatially resolved UV–optical colour maps and NUV–r profiles to trace the nature of UV emission and age of stellar population properties across the bar, bulge, and disc. As the same methodology was used in RD25, we will compare our present results with previous findings to investigate the bar evolutionary stages. If signatures of residual star formation in the nuclear region are present, it could reflect a pre-quenching phase, due to the action of the bar, during which the stellar populations in the bar region are expected to be comparatively younger than those observed for the sample in RD25.

\indent The structure of the paper is as follows. Section 2 explains the sample selection. Section 3 describes the data employed and the methodology for the analysis. Section 4 details the emission line analysis, the spatially resolved colour-colour maps to distinguish quench and unquenched morphological components for different galaxies, the radial NUV-r colour profiles and the associated dominant stellar population ages. In section 5, we discuss our results and the different evolutionary quenching mechanisms involved, and in section 6, we summarize the work. We adopt standard flat comological model with $ \mathrm{H_{0}} = 70\; \mathrm{Km \;s^{-1}\; Mpc^{-1}}, \Omega_{M}= 0.3  \; \text{and} \; \Omega_{\Lambda} = 0.7$.

\section{Sample selection}
\label{2}
The sample studied in this work consists of the most probable centrally quenched barred galaxies in the redshift range (0.01-0.06), satisfying all the initial selection criteria given in Table 1 of RD25, but showing residual or weak emission features in the 3" fibre SDSS spectra, which were obtained during the sample selection in RD25. In the sample studied by RD25, the SDSS fibre spectra of central regions showed no emission lines in general and no Halpha emission line in particular, signifying no ongoing star formation in the sub-kpc nuclear region. 
A detailed description of selecting the most probable candidates of centrally quenched barred galaxies by leveraging the differences between two SFR estimates from the  MPA-JHU and the GSWLC catalogues is provided in Section 2 and Table 1 of RD25. Here we briefly describe the criteria. In RD25, the most probable candidates for centrally quenched barred galaxies are defined as those barred galaxies that appear passive when using MPA-JHU SFR estimates but remain actively star-forming according to SFR estimates from the GSWLC catalogue (\citealt{cortese2020xgass,george2021bar}). The differences in the SFR estimates arise from the inadequate aperture corrections applied in the MPA-JHU catalogue to measure star formation outside the central 3" SDSS fibre. This leads to the classification of galaxies with extended star-forming discs and passive inner regions as passive galaxies in the MPA-JHU catalogue (as seen in Figure 1). Such galaxies are of particular interest, as they may be undergoing internal quenching processes, potentially associated with mechanisms such as AGN feedback, morphological quenching due to a central bulge (classical bulges, formed during mergers), or the influence of a stellar bar. To disentangle the effects of bar from other effects, RD25 focused exclusively on non-AGN barred galaxies that host pseudo-bulges. All systems classified as AGN or low-S/N LINERs (BPT $\leq$ 3) based on the standard BPT diagram (Baldwin, Phillips, and Terlevich; \citep{1981PASP...93....5B}) as given in the MPA-JHU catalogue, were removed, and also ensured that none of the sample galaxies host classical bulges (Sersic indices $>$2). The sersic indices for each component (bulge, bar, and disc) were provided in \cite{kruk2018galaxy}, along with detailed morphological decompositions for SDSS barred galaxies.

\begin{table*}[ht!]
\caption{Galaxies in our sample}
\label{tab:my-table}
\centering
\resizebox{0.98\textwidth}{!}{%
\begin{tabular}{llllllllllllll}
\hline \hline \\
\begin{tabular}[c]{@{}l@{}}Galaxy\\ Number\end{tabular} & \begin{tabular}[c]{@{}l@{}}RA\\ \\ (deg)\end{tabular} & \begin{tabular}[c]{@{}l@{}}DEC\\ \\ (deg)\end{tabular} & \begin{tabular}[c]{@{}l@{}}Redshift\\ \\ z\end{tabular} & \begin{tabular}[c]{@{}l@{}}Scale\tablefootmark{a}\\ \\ (kpc/arcsec)\end{tabular} & \begin{tabular}[c]{@{}l@{}}R$_{25}$\tablefootmark{b}\\ \\ (kpc)\end{tabular} &  b/a\tablefootmark{c}  & \begin{tabular}[c]{@{}l@{}}SDSS 3" \\ fiber \\ physical \\length \\ (kpc)\end{tabular} & \begin{tabular}[c]{@{}l@{}}Stellar Mass\tablefootmark{d}\\ \\ $log_{10}M_{*}$\end{tabular} & \begin{tabular}[c]{@{}l@{}}HI Mass\tablefootmark{e}\\ \\ $log_{10}M_{HI}$\end{tabular} &\begin{tabular}[c]{@{}l@{}}Bar \\ half-length\tablefootmark{f}\\ \\ (kpc)\end{tabular} & \begin{tabular}[c]{@{}l@{}}Bulge \\Sersic\\ Index\tablefootmark{g}\\ \\ n\end{tabular} &  \begin{tabular}[c]{@{}l@{}} H$\alpha$ flux\tablefootmark{h} \\\\ ($10^{-17}$ erg \\ s$^{-1}$ cm$^{-2}$$ \AA^{-1}$)\end{tabular} &   \begin{tabular}[c]{@{}l@{}} H$\alpha$ flux err\tablefootmark{i} \\ \\ ($10^{-17}$ erg \\ s$^{-1}$ cm$^{-2}$$ \AA^{-1}$) \end{tabular}\\ \\ \hline \\
1      & 352.96326 & -0.96398 & 0.0583   & 1.129 & 25.83 & 0.62 & 3.387 &11.39       &   -      & 8.05    &  0.5   & 382.3253   & 11.6073    \\\vspace{0.1cm}
2      & 242.19026 & 09.04962  & 0.0472   & 0.926 & 9.14 & 0.64 & 2.778 &10.42       &   -      & 3.67   &  0.22   &  54.4188     & 3.6728       \\\vspace{0.1cm}
3      & 262.49722 & 60.35028 & 0.0205   & 0.415 & 15.88 & 0.59 & 1.245 &10.87       &   -      & 3.45    &   0.67  &  1418.1600    & 26.547    \\\vspace{0.1cm}
4      & 176.48159 & 21.02552 & 0.0225   & 0.455 & 19.08 & 0.85 & 1.365 &10.86       &   9.89      &  5.77    &  1.01  &  174.6633  &  7.5108    \\\vspace{0.1cm}
5      & 213.89783 & 03.92578  & 0.0556   & 1.080  & 20.94 & 0.87 & 3.240 &11.04       &   -      & 7.27   &   0.77  & 243.4204  &  7.3642        \\\vspace{0.1cm}
6      & 327.00214 & -0.97734  & 0.0522   & 1.018 & 11.67 & 0.73 & 3.054 &10.65       &   -      & 4.43   &   2.82   &  203.6573   &  2.5036     \\\vspace{0.1cm}
7      & 222.21018 & 57.77085 & 0.0435   & 0.857 & 11.88 & 0.77 & 2.571 &11.05        &   
-      &  6.79    &   0.94  &  616.5107    &  26.988     \\\vspace{0.1cm}
8      & 219.07022 & -1.38696 & 0.0556   & 1.080  & 19.65 & 0.90  & 3.240 &11.14       &   -      &  10.26  &   0.91 & 139.6327    &  4.3751       \\\vspace{0.1cm}
9      & 153.76113 & 05.67692  & 0.0287   & 0.576 & 10.96 & 0.81 & 1.728 &10.55       &   -      &  4.56    &  1.13   & 359.8664   & 11.6276     \\\vspace{0.1cm}
10     & 340.49181 & -0.78654  & 0.0533   & 1.038 & 15.94 & 0.79 & 3.114 &10.31       &   -      &  3.44   & - & -    & -   \\\vspace{0.1cm}
11     & 228.37500   & 02.10417  & 0.0385   & 0.763 & 10.96 & 0.77 & 2.289 &10.56       &   -      & 3.63    &    0.46  & -    & -  \\\vspace{0.1cm}
12     & 213.56730  & 10.58274 & 0.0320    & 0.639 & 9.36 & 0.52 & 1.917 &10.96       &  -       & 4.55   &   1.21    & 513.6786    & 11.5014    \\
\hline
\label{1}
\end{tabular}
}
\tablefoot{
\tablefoottext{a}{Angular to physical scale length}
\tablefoottext{b}{R$_{25}$ scale length taken from NED.}
\tablefoottext{c}{Major-to-minor axis ratio}
\tablefoottext{d}{Stellar mass taken from GSWLC M2 catalogue}
\tablefoottext{e}{HI mass obtained from the ALFALFA-SDSS galaxy catalogue \citep{2020AJ....160..271D}}
\tablefoottext{f}{Bar half-length estimated from our analysis}
\tablefoottext{g}{Bulge Sersic Index derived from \cite{kruk2018galaxy}
\tablefoottext{h,i}H$\alpha$ fluxes with corresponding errors estimated from Gaussian fit to continuum subtracted emission spectra.}

}

\end{table*}

\indent Interestingly, during the construction of the RD25 sample, we came across a set of 12 barred galaxies (not included in RD25) that exhibited central emission in their SDSS fibre spectra, despite satisfying all the selection criteria described above for being centrally quenched (see Figure 1). These sample galaxies could represent the final evolutionary phase in the bar-quenching sequence, where star formation is confined to the nuclear sub-kpc region, while the region between the bar ends and the center is already quenched. In fact, most of these galaxies are classified as composite (indicating ionisation sources from star formation and/or AGN, and/or shocks) based on the standard BPT diagram ([NII]/H$\alpha$ diagnostic diagram) in the MPA-JHU catalogue.  In this work, we study these 12 galaxies. The details of the sample are given in Table 1. Neutral hydrogen (HI) mass measurements are available for only one galaxy, obtained from the ALFALFA–SDSS catalogue \citep{2020AJ....160..271D}. Their HI-to-stellar mass ratios ($\sim$10 \%) are consistent with values typical of massive star-forming disc galaxies. To assess their local environments, we constructed 1 Mpc cubic volumes centered on each galaxy and identified neighbors using the NASA Extragalactic Database (NED). Roughly half of the galaxies appear isolated or have only one close neighbor, while a few lie in moderately populated regions. In particular, Galaxy 4 has nine neighbors, while Galaxies 1, 10, and 11, each have three neighbors. However, previous studies have shown that for galaxies with stellar masses above $10^{10.5} M_{\odot}$, environmental effects on star formation are generally negligible \citep{2010A&A...509A..40I, 2010ApJ...721..193P, Guo_2019}. These findings suggest that the observed central emission is unlikely to be environmentally driven. 

\begin{figure}
\centering
\begin{subfigure}{0.45\textwidth}
   \includegraphics[width=\linewidth]{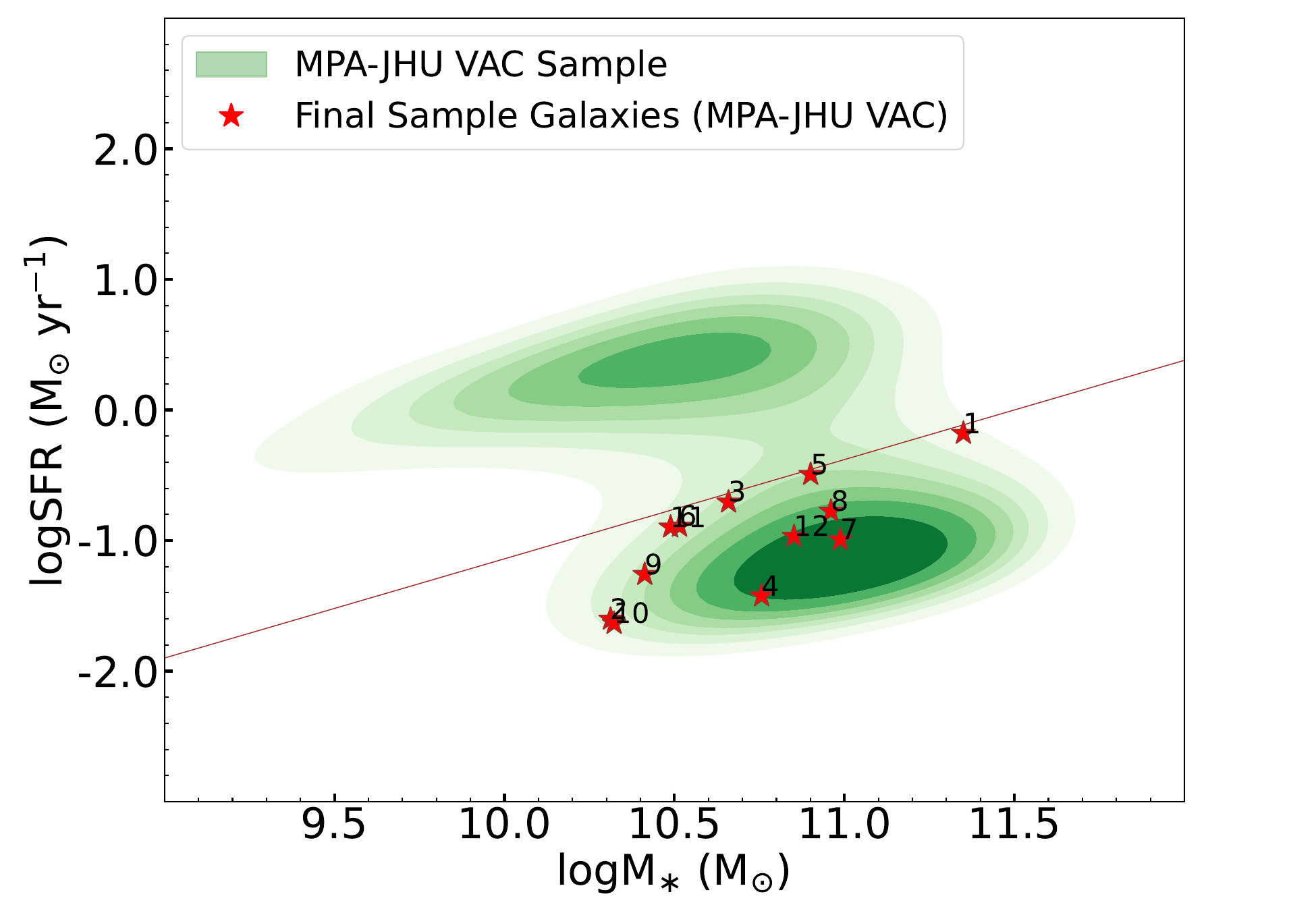}
   \subcaption{}
\end{subfigure}
\par\medskip
\begin{subfigure}{0.45\textwidth}
   \includegraphics[width=\linewidth]{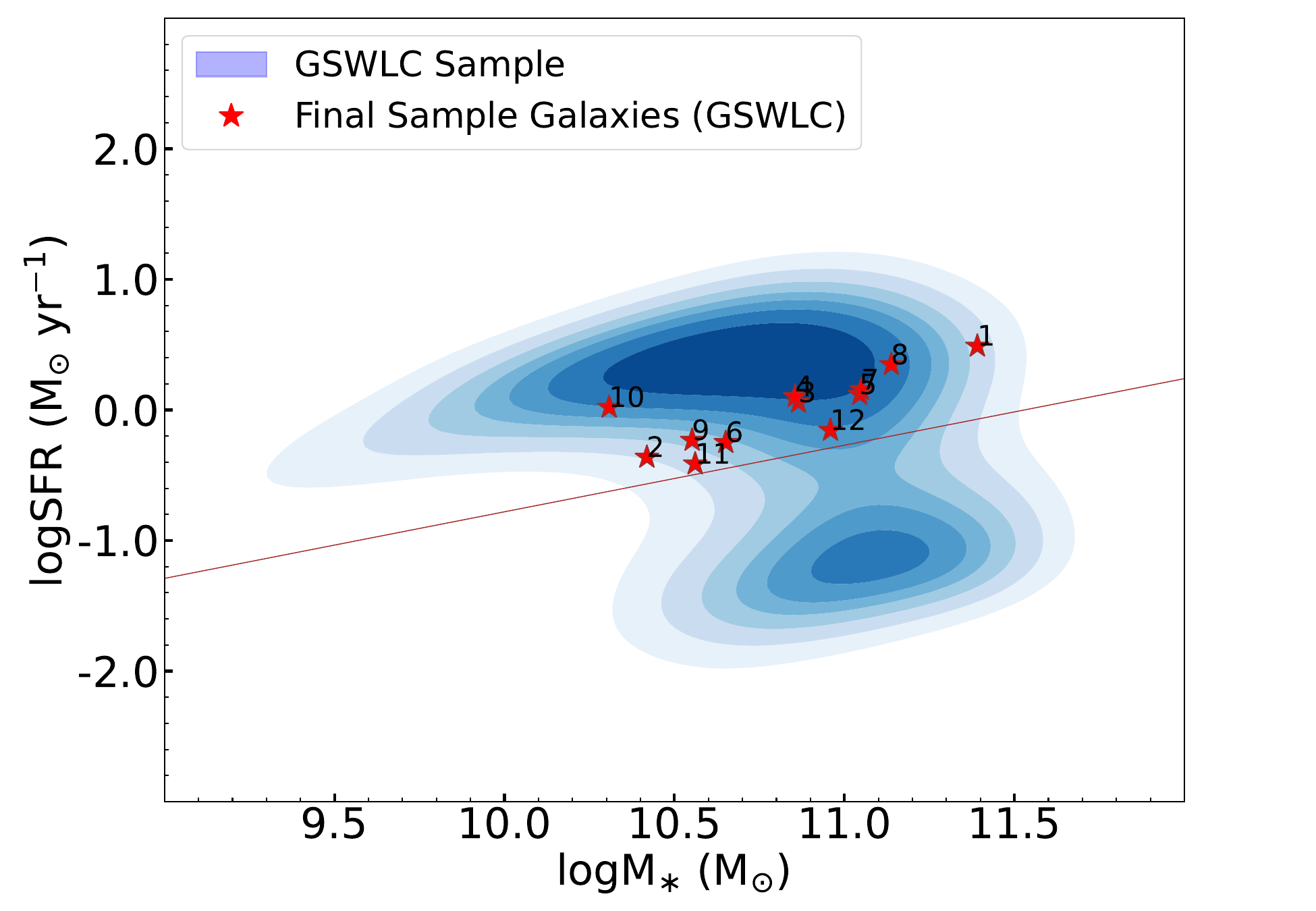}
   \subcaption{}
\end{subfigure}
\caption{ SFR-M$_{*}$ relation defined for the final sample of 12 barred galaxies with two different SFR indicators: MPA-JHU Value Added Catalogue and GSWLC M2 catalogue. (a) SFR-M$_{*}$ relation for MPA-JHU VAC \citep{brinchmann2004physical} (green contours) with red line representing SFMS$-$1.1 dex adopted from \cite{2020MNRAS.499..230B}. (b) SFR-M$_{*}$ relation for GSWLC catalogue \citep{2018ApJ...859...11S} (blue contours) with red line representing SFMS$-$0.6 dex taken from \cite{Guo_2019}. The -0.6 dex cut here is higher than the conventional threshold used to define passive galaxies. Selecting galaxies above this cut are preferentially identified as non-passive systems.}
\label{fig1}
\end{figure}

\begin{figure*}[h!]
\centering
    \includegraphics[width=0.9\linewidth]{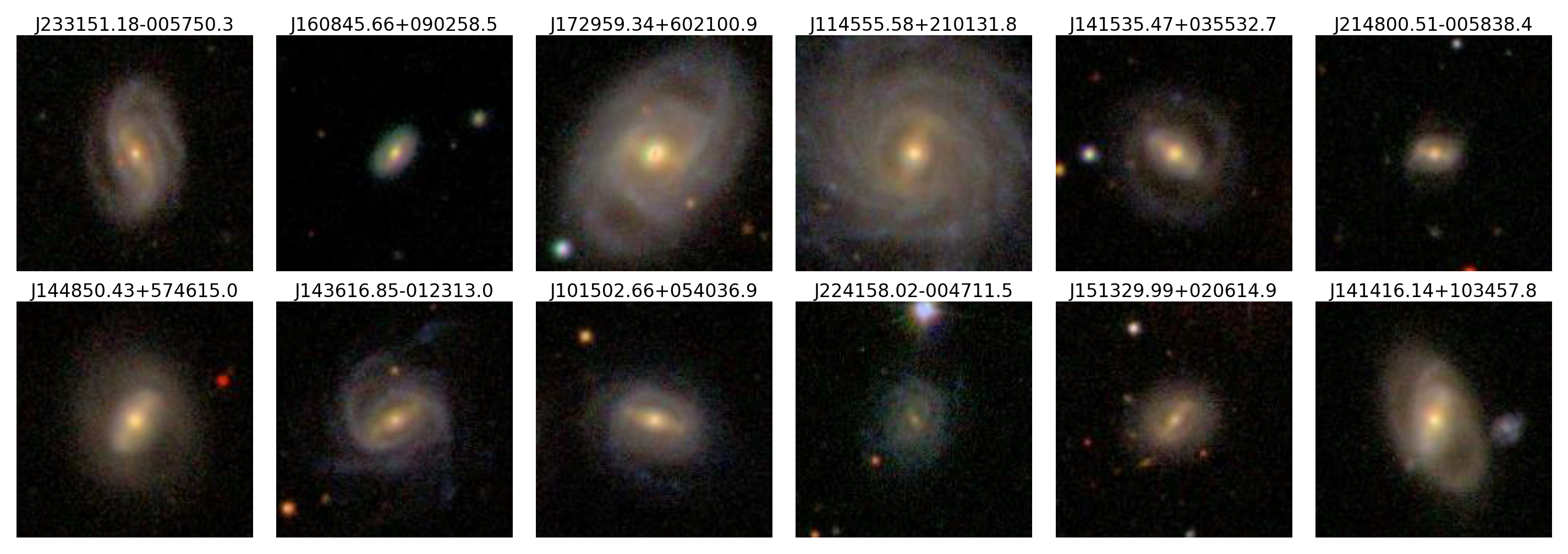}\
    \caption{Postage stamp of colour composite images of the final sample for {\bf 12} barred galaxies from SDSS DR7 u,g,r,i, and z bands. Each image is $\sim$ 1 arcmin $\times$ 1 arcmin with redshift ranging from 0.01- 0.06. The SDSS id is also provided on the top of each image. More information on the sources is available in Table 1.}
    \label{fig2}
\end{figure*}

\section{Data and Methodology}
\label{3}
The sample (Figure 2) lies in the redshift range 0.01-0.06 and includes galaxies with an inclination cut of less than 60° (i.e., axis ratio b/a > 0.5). All galaxies have been observed in either medium, deep, or GII imaging surveys from GALEX, with exposure times exceeding 1000 seconds. The galaxies have stellar masses greater than $M_{\ast} > 10^{10}\ M_{\odot}$. Our first aim is to  better understand the source of ionisation in these systems using emission line analysis and [SII] based BPT diagram. We then study the spatially resolved UV-optical colour maps and NUV-r radial profiles to understand the nature of UV emission and the age of stellar populations in different regions of the galaxies. We use the imaging and spectroscopic data as described below.

\indent As mentioned in Section 2, according to the MPA-JHU catalogue, the nature of ionizing sources producing these emission lines in these galaxies is classified as composite (star formation/AGN/shocks). The BPT classification in the MPA-JHU catalogue was made using [NII]/Hα-based diagnostic diagram. To further characterize the nature of the ionizing source, we additionally adopt the [SII]-based BPT diagnostic in our study. It enables a separation between Seyfert-like and LINER-like excitation among galaxies classified as AGN or composite in the [NII]/Hα diagnostic diagram \citep{kauffmann2003stellar, 2006MNRAS.372..961K}. We use the Sloan Digital Sky Survey (SDSS) latest Data Release (DR18, \cite{2023ApJS..267...44A}) spectra (covering the central 3" region of sample galaxies) to analyze the emission properties of our sample. The spectra are obtained using the dedicated 2.5-meter Sloan Foundation Telescope equipped with a 640-fiber double spectrograph at Apache Point Observatory, New Mexico. Each fiber subtends 3 arcseconds on the sky and feeds the light into a dual-arm spectrograph that covers a wavelength range of approximately 3800–9200 \AA\ with a spectral resolution of R $\approx$ 1800--2200. The spectra are already rebinned in the wavelength domain and have signal-to-noise ratios (S/N) exceeding 20 for most objects except for two galaxies - galaxy 10 (S/N $\approx$ 5) and galaxy 11 (S/N $\approx$ 15). We then apply full spectral fitting to each spectrum using the pPXF (Penalized Pixel-Fitting) method \citep{2004PASP..116..138C, 2017MNRAS.466..798C}, which enables simultaneous modelling of the stellar continuum and emission lines and extracting key emission line properties from each spectrum using BPT diagnostics based on both [NII] and [SII] emission lines. This method facilitates accurate measurements of weak or blended lines and improves the deblending of H$\alpha$ and [NII]. The details of the emission line analysis and BPT classification are given in Section 4.\\
\indent Our sample galaxies are those that are classified as quenched based on SFR estimates from the MPA-JHU catalogue, but are star-forming based on SFR estimates from the GSWLC. This suggests that these galaxies have recent star formation in their extended discs and are passive in the inner regions. Although the UV emission in galaxies is predominantly from young star-forming regions (which host massive O and B-type stars) evolved, hot stellar populations such as the
horizontal branch stars, post asymptotic giant branch stars and white dwarfs can emit in UV and this excess UV emission (in FUV) due to old and hot (or extreme) horizontal branch stars is known as UV upturn. The UV-optical colour-colour map, provided by \cite{2011ApJS..195...22Y}, makes use of two colour indices, FUV$-$NUV and NUV$-$r, to understand the nature of UV emission in galaxies and classify them into UV upturn, UV weak or star-forming categories. As we did in RD25, we analyse the UV-optical colour maps of our 12 sample galaxies.\\
\begin{figure}[h]
   \centering
    \includegraphics[width=0.75\linewidth]{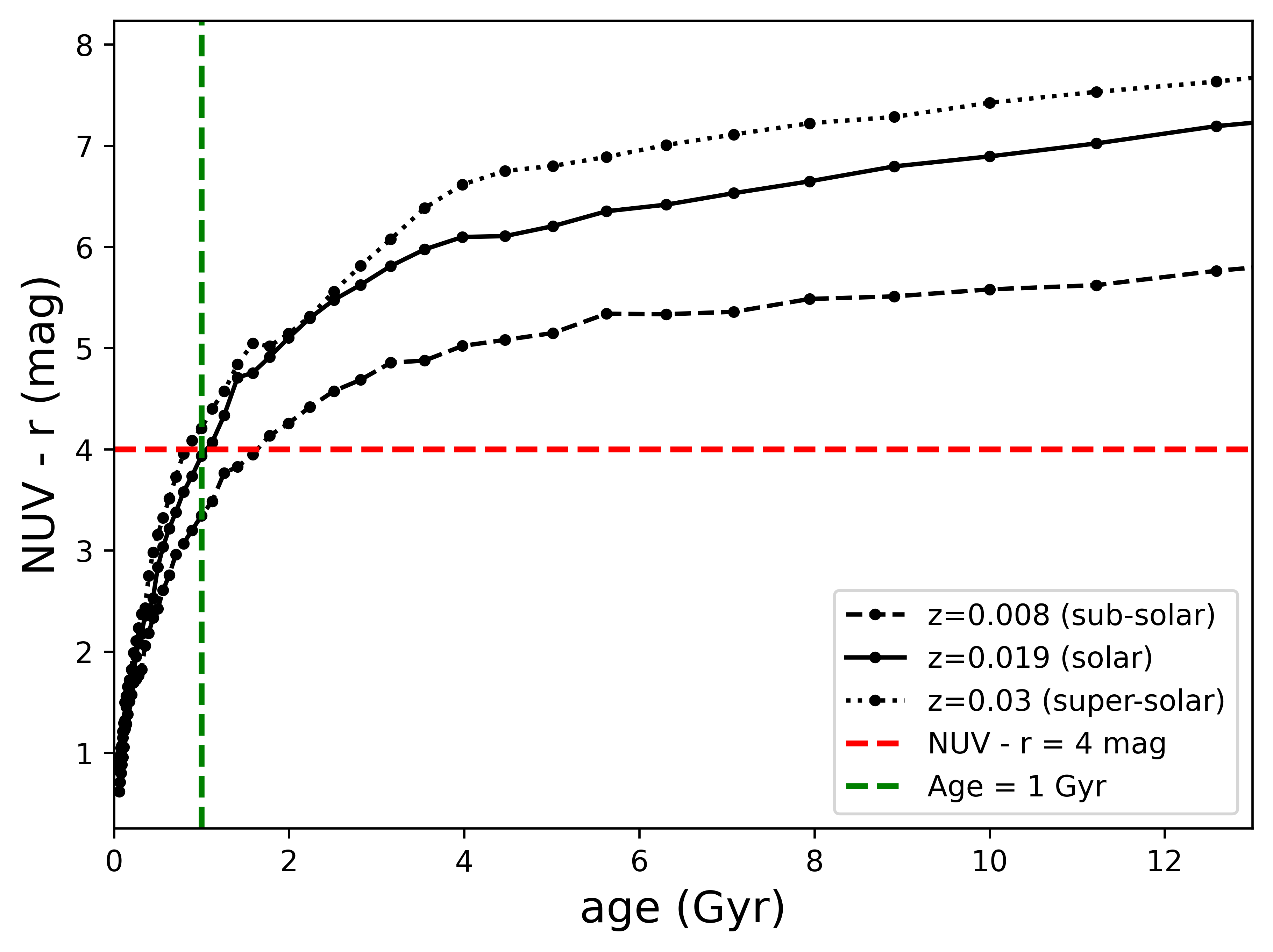}
\caption{ The NUV–r colour as a function of stellar age by using photometric predictions based on the E-MILES stellar population synthesis models.}
\label{3}
\end{figure}
\indent To constrain the age of the stellar population in the different regions of our sample galaxies, we analyse the radial profiles of NUV$-$r colour, which is an excellent photometric indicator of star formation in the last 1 Gyr \citep{2005ApJ...619L..39S, 2007ApJS..173..619K, 2014SerAJ.189....1S, Pan_2016}, of our sample galaxies. The NUV–r colour is known to correlate with the Dn4000 index, with a typical NUV-r colour of $\sim$ 4 mag corresponding to a Dn4000 value of 1.5 \citep{Pan_2016, kauffmann2003stellar}, which is often associated with post-starburst or recently quenched stellar populations. This supports the use of NUV–r as an effective age indicator for stellar populations up to $\sim$ 1–2 Gyr. However, to ensure consistency with these findings and with our analysis framework, we have estimated the expected NUV–r colours as a function of stellar age by using photometric predictions based on the E-MILES (Extended Medium-resolution Isaac Newton Telescope Library of Empirical Spectra{\footnote{\url{https://research.iac.es/proyecto/miles/}}}, \citealt{2016MNRAS.463.3409V}) stellar population synthesis models. The E-MILES library incorporates multiple high-quality stellar libraries: the NGSL space-based library in the UV range (1680–3540 \AA\ ), the MILES stellar library in the optical range (3525–7500 \AA\ ), and the IRTF spectral library in the near-infrared range (0.8–5.2 $\mu$m). These libraries are combined to provide broad wavelength coverage and are used to produce grids of stellar population synthesis (SPS) models that span a range of parameters, including age, metallicity, and initial mass function (IMF). The model SEDs are generated assuming a revised Kroupa IMF with a slope of 1.3, covering a metallicity (Z) range from 0.0001 to 0.03 and an age range of 63 Myr to 17 Gyr, based on the Padova+00 isochrones \citep{2016MNRAS.463.3409V}. Then the photometric predictions are derived by convolving theoretical spectral energy distributions (SEDs) with the exact SDSS and GALEX filter transmission curves. This helps us obtain a self-consistent age–colour relation tailored to our sample, providing a more reliable comparison for interpreting observed colour profiles. As shown in Figure 3, the synthetic NUV–r colour increases with stellar age for all metallicities, with redder colours predicted at higher metallicities. For solar metallicity (Z = 0.019), an age of approximately 1 Gyr corresponds to NUV–r $\approx$ 3.96 mag, consistent with the empirical relation reported by \citet{Kaviraj_2007} and also with the threshold adopted in RD25. This strengthens the reliability of using NUV–r colour as a photometric proxy for tracing recent star formation history within the past 1–2 Gyr in our sample galaxies.\\
\indent For the photometric analysis and to construct spatially resolved UV-optical colour maps, we use r-band photometric data from SDSS DR7 (\citealt{2009ApJS..182..543A}), which have a pixel scale of 0.396" and a typical PSF of 1.32". We extract 2' $\times$ 2' image cutouts centered on each galaxy and perform background subtraction using mean values from source-free regions. Foreground and nearby sources are masked. We also use imaging data from the Galaxy Evolution Explorer (GALEX) in the FUV (1344-1786 \AA\; $\lambda_{\text{eff}}$ = 1538.6 \AA\ ) and NUV (1771-2831 \AA\; $\lambda_{\text{eff}}$ = 2315.7 \AA\ ) bands, accessed through the MAST portal \citep{2005ApJ...619L...1M}. The full width at half maximum (FWHM) of the FUV and NUV channels is 4.2" and 5.3", respectively \citep{2007ApJS..173..682M}. To ensure consistent spatial resolution, the SDSS r-band and GALEX FUV images are convolved to match the NUV PSF using 2D Gaussian kernels derived from their respective point spread functions (PSFs). The kernel widths are computed using the equation $\sigma_{K} = \sqrt{(\sigma_{target})^{2}-(\sigma_{original})^{2}}$ and the kernels are normalised to keep the flux conserved after convolution. We adopt kernel widths, $\sigma$ = 2.19 arcsec for the r-band and $\sigma$ = 1.38 arcsec for the FUV image. These PSF-matched, background-subtracted, and masked images are then used to construct spatially resolved colour maps and NUV–r radial profiles. However, native-resolution r-band images are retained for measuring bar lengths. The bar lengths are measured using the photutils ellipse-fitting routine applied to the original (non-degraded) SDSS r-band images, following the methodology described in RD25. The resulting bar lengths for individual galaxies are listed in Table 1.It is essential to note that for some galaxies, the signal-to-noise ratio (S/N) in the outer regions of the GALEX NUV images is quite low, which affects the reliability of the derived NUV–r colour. Therefore, we apply a threshold and exclude pixels where the signal-to-noise ratio (S/N) falls below 3, as the colour measurements in these regions are not physically meaningful due to noise blending.

\section{Analysis}
\label{section4}
\begin{figure*}[h!]   
   \centering
    \includegraphics[width=1.0\linewidth]{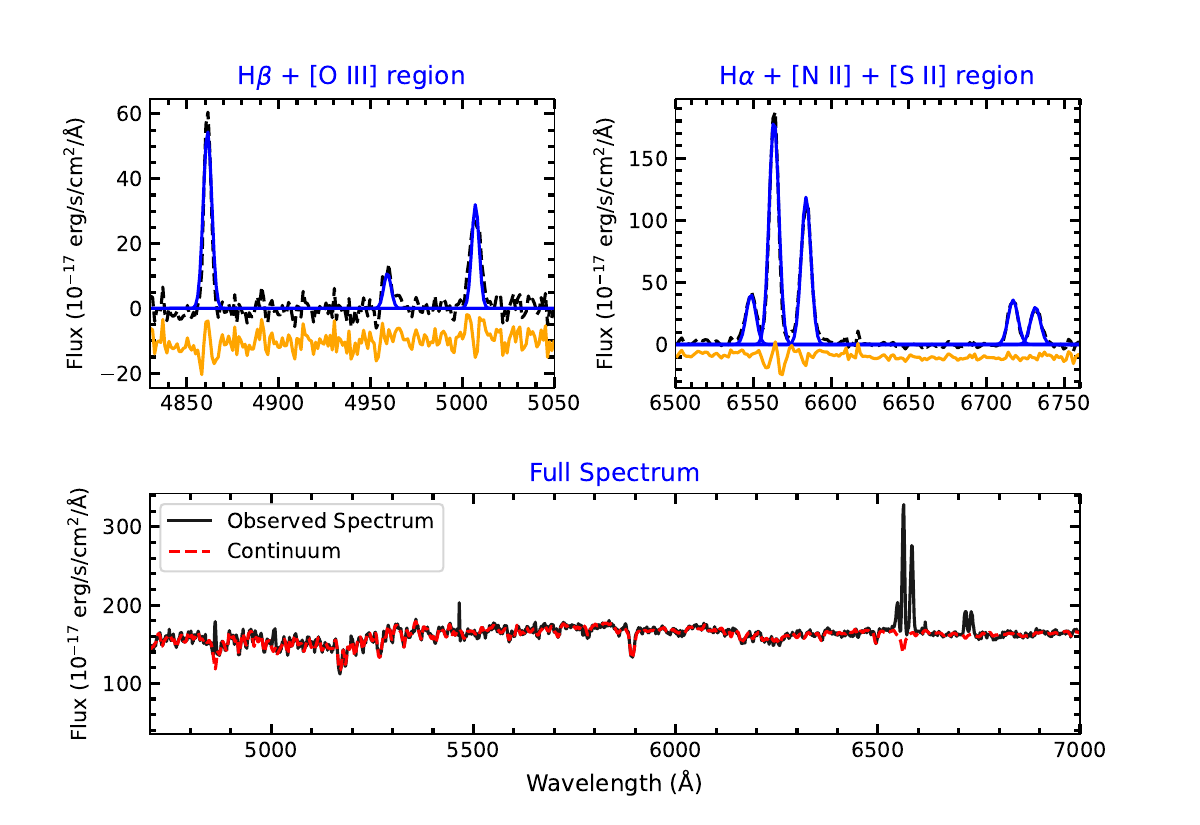}
\caption{The pPXF fit is shown to the observed spectrum in black, with the best-fit optimal continuum overlaid in red. Upper panels: The emission-line spectrum, obtained after subtracting the continuum, is shown in black. The Gaussian fits to the emission lines are displayed in blue, and the residuals are plotted in orange.}
\label{fig3}
\end{figure*}
\subsection{Emission Line Analysis}
We analysed the observed SDSS DR18 spectra of 12 barred galaxies to investigate the nature of the ionising source that produces Hα emission. First, we corrected for foreground dust extinction using the Milky Way-type extinction law from \cite{1989ApJ...345..245C}, applying E(B-V) values from \cite{2011ApJ...737..103S}. To extract emission line ratios, we employed the pPXF code, which models the underlying stellar population by fitting a linear combination of spectral templates to the observed spectrum while convolving it with the line-of-sight velocity distribution (LOSVD). Since pPXF operates in pixel space, it allows for the masking of bad pixels and the exclusion of gas emission lines during continuum fitting.
For full-spectrum fitting, we used the EMILES stellar population synthesis (SPS) model templates by \cite{2016MNRAS.463.3409V}, which are based on Padova isochrones and assume a Salpeter initial mass function (IMF). These spectral templates span a wavelength range of 3500 \AA\ to 10,000 \AA\, and for each galaxy we used the portion of the observed rest-frame spectrum that falls within this range for template matching, which we adjusted to match the rest-frame observed range. The input flux values were normalised by dividing by the median flux, and initial guesses for velocity (linked to redshift) and velocity dispersion were provided. pPXF employs a Gaussian-Hermite parameterisation for the LOSVD with the number of moments set to four, allowing simultaneous determination of velocity, velocity dispersion ($\sigma$), higher order moments $h_{3}$ and $h_{4}$. pPXF then generates a best-fit model spectrum by convolving the template spectrum with the LOSVD. The optimal parameters are determined through nonlinear least squares optimisation. The reddening-corrected spectra and corresponding best-fit models for one galaxy are shown in Figure 4. The pure emission-line spectrum is obtained by subtracting the best-fit model from the reddening-corrected spectra.\\
\indent For profile-fitting of emission lines, we use the standard method described in \cite{1997ApJ...487..568H} and later a similar method adopted in \citep{2004ApJ...610..722G, 2013ApJ...775..116R} and \cite{2016MNRAS.455.3148S}. The line profile of [SII] $\lambda\lambda$6716,6731 doublet is empirically assumed to be similar to [NII] $\lambda\lambda$6548,6583 doublet and narrow H$\alpha$ lines. The [SII] doublet is first fitted with a combination of two Gaussian components with the assumption that the widths of the two lines in velocity space are equal and the relative separation between the two lines is fixed by their laboratory wavelengths. Then, using the template of [SII] doublet, we fit [NII] + H$\alpha$ narrow lines modelled with three Gaussian components, assuming equal width to [SII] doublet \citep{1988ApJ...324..134F, 1989ApJ...342L..11F, 1997ApJS..112..391H}. The relative separation between [NII] lines is also fixed by their laboratory wavelengths, and their relative strengths are fixed with a theoretical value of 2.96. The H$\beta$ line, as well, is fitted with an equal width assumed for narrow lines \citep{2013ApJ...775..116R}. Except we fit the [OIII] lines with no constraints since the O[III] lines get affected by outflows, resulting in an asymmetric profile with a broad blue shoulder \citep{1986ApJ...301...98D, 1985MNRAS.213...33W}. After fitting, the emission fluxes are calculated using Gaussian model parameters. As mentioned in Section 3, the S/N of the spectra of objects 10 and 11 is low (5 and 15), and hence the spectra fits were not satisfactory, and the very weak emission lines in their spectra were not identified or fit well. We classify them as no-emission line galaxies and exclude them from further analysis. However, later, during the comparison with the RD25 sample, we included them with the RD25 sample. 

\subsection{Source of ionisation: BPT classification}
\label{subsection4.2}
After deriving the emission-line fluxes and corresponding flux ratios, we determine the location of our sample galaxies in both BPT diagnostic diagrams ([NII]/H$\alpha$ vs [OIII]/H$\beta$, hereafter BPT-1, and [SII]/H$\alpha$ vs [OIII]/H$\beta$, hereafter BPT-2) as shown in Figure 4. Here, the use of the [SII]-based diagnostic (BPT-2) is particularly helpful for disentangling the dominant ionisation mechanisms in sources classified as composite in BPT-I. We note that [O I]-based BPT diagnostic diagrams can also be used to understand the dominant emission mechanism in systems classified as composite in BPT-I. However, the [O I]-based BPT diagnostic is not considered here because [O I] lines are too weak to detect in the spectra of our sample galaxies. We therefore restricted our analysis to BPT-1 and BPT-2 diagnostic diagrams, which are based on emission lines with robust measurements from the spectra of our sample galaxies. In BPT-1, seven objects are classified as composite and three as AGN. Objects 10 and 11 are not shown because their spectra could not be properly fitted due to their low signal-to-noise ratio. In BPT-2, among the six composite objects from BPT-1, objects 1, 3, 5, 6, 7, and 12 are further classified as star-forming, while object 9 falls into the LINER category. Of the three AGN-classified objects in BPT-1, objects 2, 4, and 8 are identified as LINERs in BPT-2. We next explore the underlying ionisation sources that may account for the observed LINER-like emission characteristics.

\begin{figure}[h!]   
\centering
    \begin{subfigure}{0.45\textwidth}
   \includegraphics[width=\linewidth]{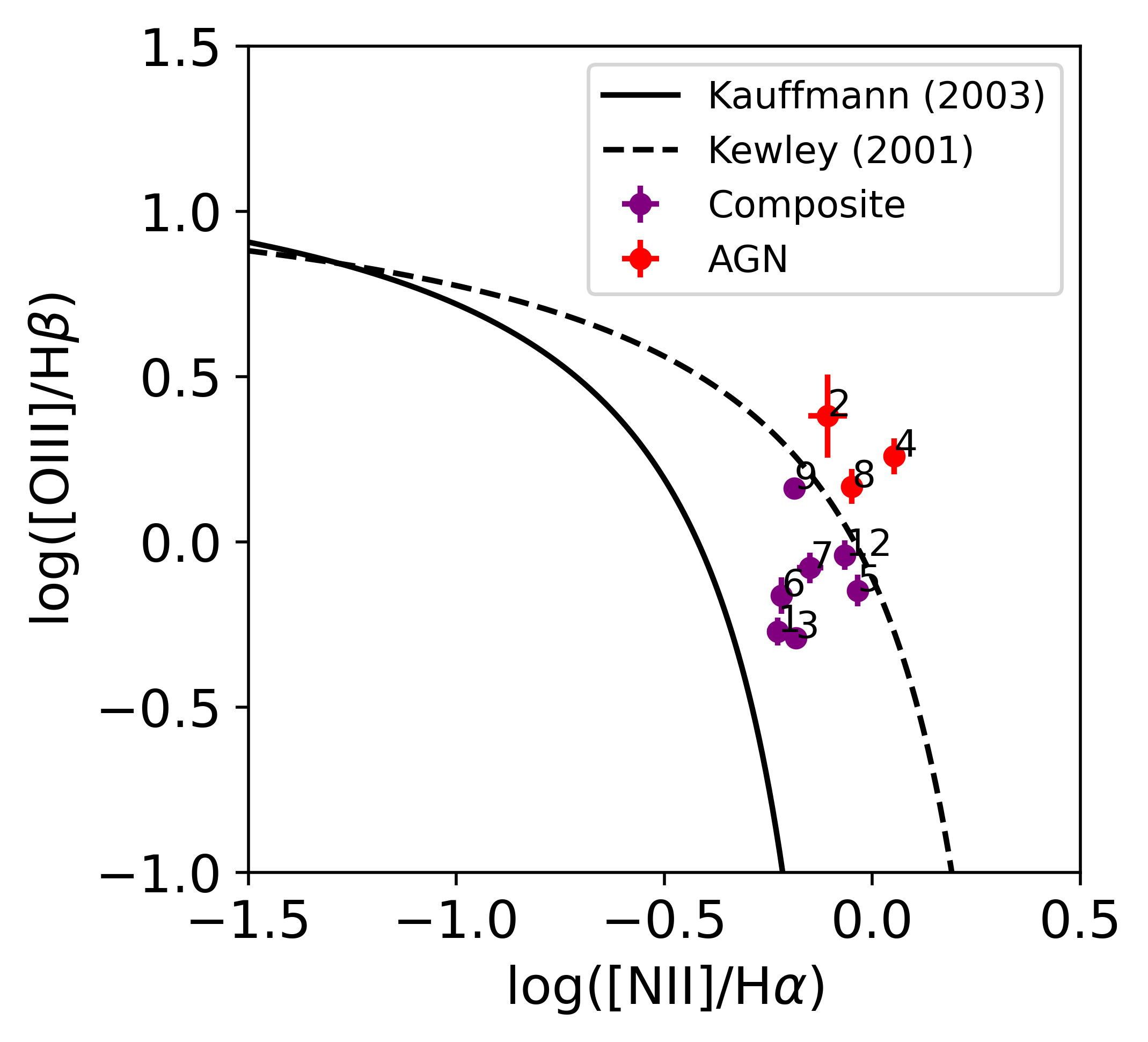}
   \subcaption{}
\end{subfigure}
\par\medskip
\begin{subfigure}{0.45\textwidth}
   \includegraphics[width=\linewidth]{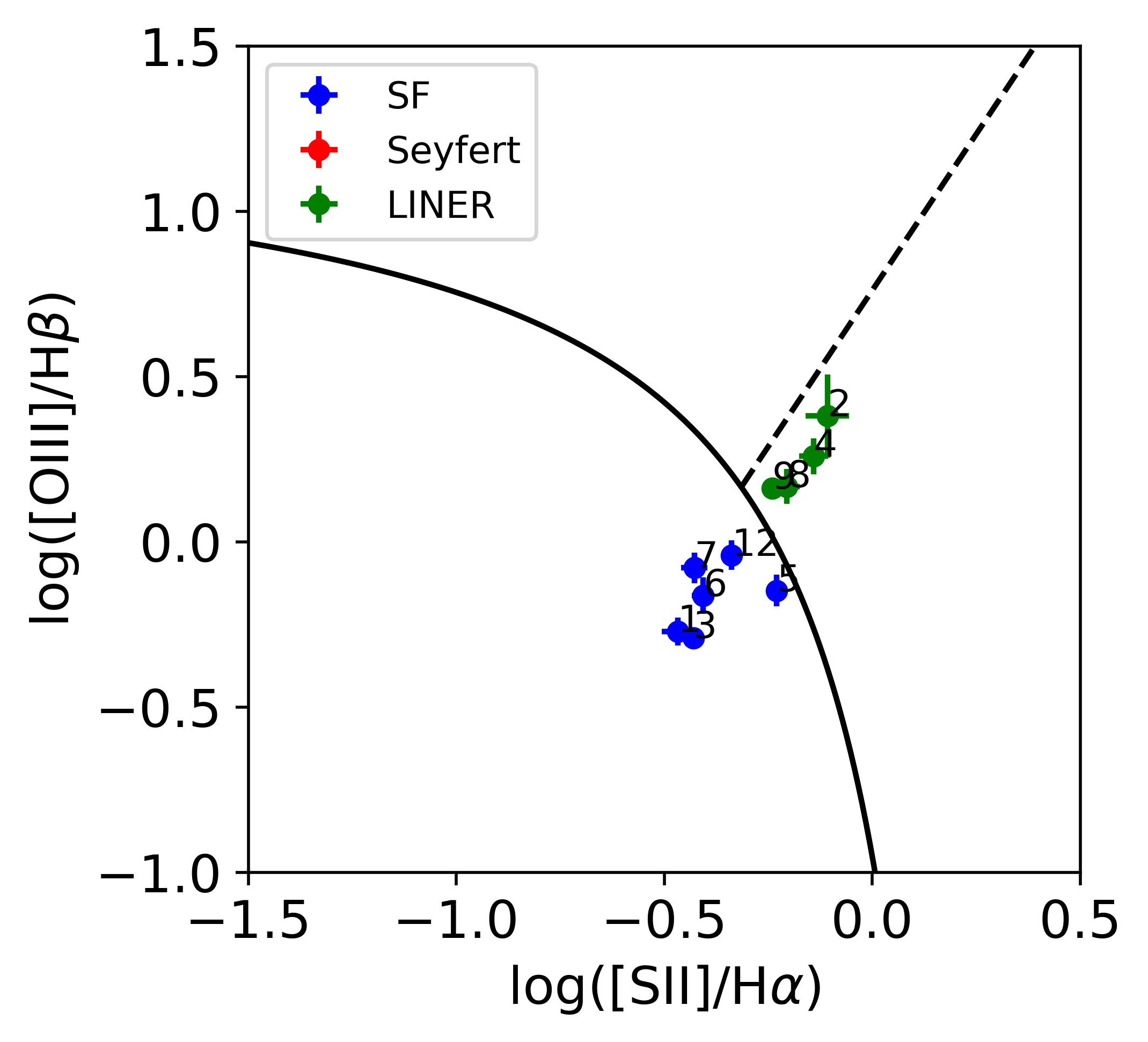}
   \subcaption{}
\end{subfigure}

\caption{(a) The primary BPT diagnostic diagram (BPT-1), [NII]/H$\alpha$ vs [OIII]/H$\beta$. The dashed and solid lines represent demarcation lines taken from \cite{kauffmann2003stellar} and \cite{2001ApJ...556..121K}. These lines divide the galaxies between star-forming, composite, and AGN systems. (b) The secondary BPT diagnostic diagram (BPT-2), ([SII]/H$\alpha$ vs [OIII]/H$\beta$. The dashed line here is the demarcation line from \cite{2006MNRAS.372..961K}, and it is used to further classify galaxies into Seyferts and LINER systems.}
\label{fig4}
\end{figure}

\begin{figure*}[h!]   
   \centering
    \includegraphics[width=0.9\linewidth]{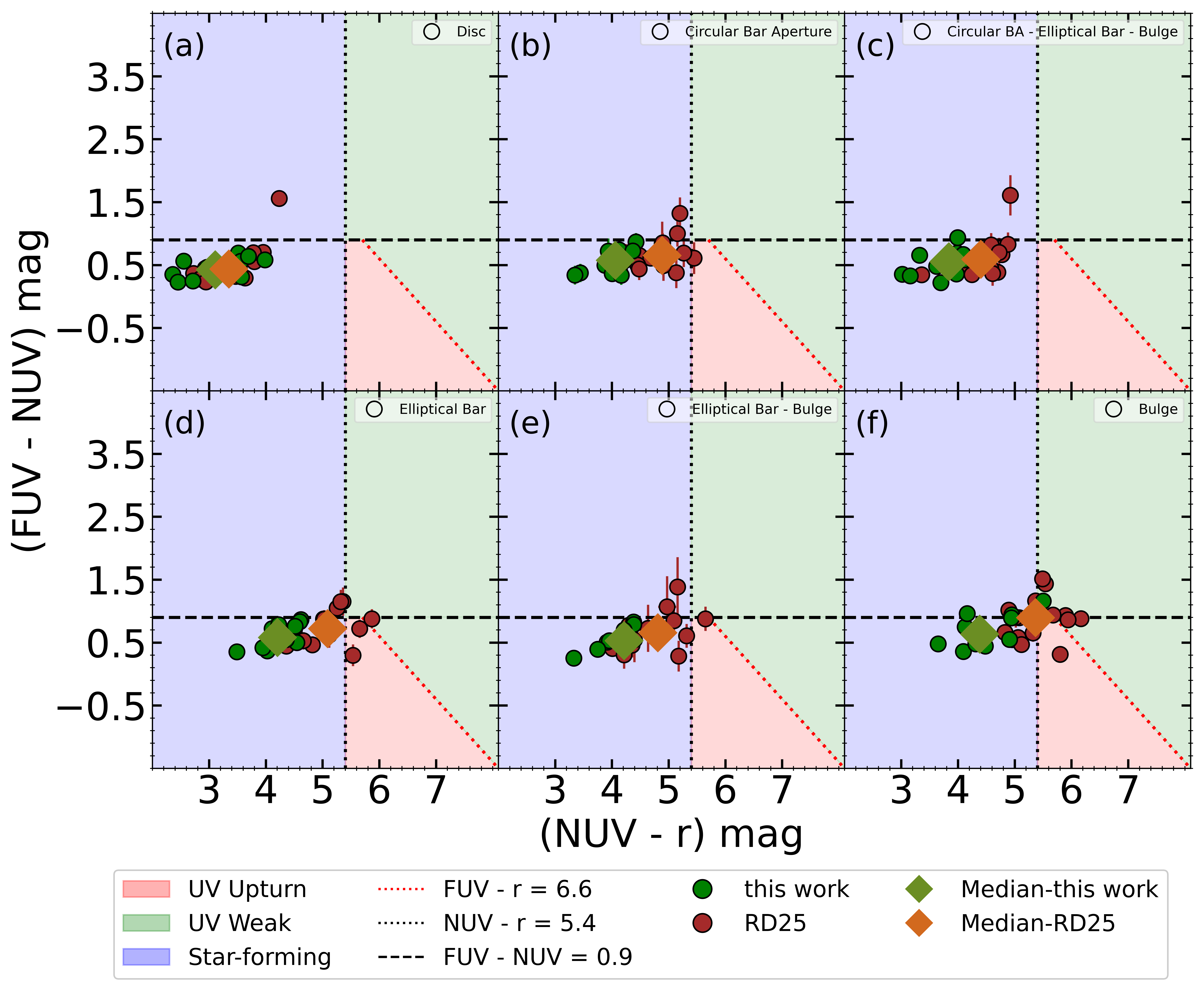}
\caption{UV-optical spatially resolved colour-colour maps for (a) the disc component, (b) the circular bar aperture region without subtracting the bulge component, (c) the circular bar aperture after subtracting the bulge and bar components, (d) the elliptical bar aperture region without subtracting the bulge, (e) the elliptical bar aperture region after subtracting the bulge, and (f) the bulge region. The lightgreen and brown diamond symbols denote the median values of the current sample and RD25 in each map. The spatially resolved two colour maps reveal that the inner region, bar and bulge regions of current galaxies within circular bar aperture, are redder ($\geq$ 4 mag), indicating ages of the older stellar populations  $>$1 Gyr. However, when compared to the RD25 sample, the colours are relatively bluer.}
\label{fig5}
\end{figure*}

\begin{table}[h!]
\centering
\captionsetup{width=1.0\linewidth}
\caption{Estimated black hole masses for all sample galaxies.}

\resizebox{0.32\textwidth}{!}{%
\begin{tabular}{ccc}
\hline\hline \\
       \multicolumn{1}{c}{\begin{tabular}[c]{@{}c@{}} Galaxy\\ Number\\ \end{tabular}} & \multicolumn{1}{c}{\begin{tabular}[c]{@{}c@{}} Velocity\\ dispersion \\and errors \\ {[}$Kms^{-1}${]}\end{tabular}}   & \multicolumn{1}{c}{\begin{tabular}[c]{@{}c@{}} Black hole\\ mass \\and errors\\ {[}$log_{10}(M_{BH}/ M_{\odot})${]}\end{tabular}}\\ \\ \hline \\

        1 & 153.60 $\pm$ 5.44 &7.48 $\pm$ 0.03  \\\vspace{0.1cm}
        2 & 81.23 $\pm$ 5.22 &6.09 $\pm$0.06 \\\vspace{0.1cm}
        3 & 122.11 $\pm$ 3.55 &6.98 $\pm$0.03 \\\vspace{0.1cm}
        4 & 141.60 $\pm$ 4.90 &7.31 $\pm$0.03  \\\vspace{0.1cm}
        5 & 133.02 $\pm$ 4.99 &7.17 $\pm$0.03\\\vspace{0.1cm}
        6 & 96.50 $\pm$ 4.54 &6.47 $\pm$0.04 \\\vspace{0.1cm}
        7 & 173.68 $\pm$  5.32 &7.76 $\pm$0.03\\\vspace{0.1cm}
        8 & 122.70 $\pm$ 5.76 &6.99 $\pm$0.04 \\\vspace{0.1cm}
        9 & 102.80 $\pm$ 5.13 &6.61 $\pm$0.05  \\\vspace{0.1cm}
       12 & 124.73 $\pm$  4.14 &7.03 $\pm$0.03 \\
       \hline
\end{tabular}
}
\label{table2}
\end{table}

The LINER-like emission is commonly associated with low-luminosity AGN (LLAGN), where the accretion of SMBH occurs at a relatively slower rate \citep{1980A&A....87..152H, 1997ApJ...487..568H, 2008ARA&A..46..475H}. Alternative explanations for LINER activity have also been proposed in the literature. Ionisation from hot post-asymptotic giant branch (post-AGB) stars and shocks can mimic LINER-like line ratios \citep{2025ApJ...984..106N}. \cite{2013A&A...558A..43S} have demonstrated that massive galaxies with little ongoing star formation but with stellar populations older than 1 Gyr can produce LINER-like ionisation in the presence of gas. This finding is consistent with our results in RD25, which indicate that the inner regions of barred galaxies host stellar populations older than 1 Gyr, reinforcing a possible link to bar-driven quenching. Furthermore, earlier studies have shown that regions affected by strong shocks and shear often exhibit composite or LINER-like ionisation ratios \citep{2004A&A...413...73Z, 2015MNRAS.446.2468E, 2020ApJ...898..116K, Kim_2024}. These high-resolution studies have shown that the bar region is strongly influenced by shear and shocks resulting from radial gas inflows, whereas the nuclear regions are more often dominated by starburst activity. However, the physical extent of the nuclear region in such studies is typically smaller than what we probe here. In our case, the emission is measured within the SDSS fibre aperture, which spans 1.24–3.39 kpc across the redshift range of our sample galaxies. This is considerably larger than the conventional definition of a nuclear region ($<1$ kpc) in most galaxies. It may be possible that strong shock and shear could influence the observed emission; however, quantifying their exact role lies beyond the scope of this study. It is also plausible that part of the LINER-like emission extends along the bar region itself \citep{2020ApJ...898..116K}, where bar-induced shocks and turbulence may ionise the gas. Confirming this would require detailed, spatially resolved spectroscopic observations capable of mapping the ionisation structure across the bar. \\
\indent From the analysis of emission lines in the central 3" arcsec fibre spectra of our sample galaxies, we find that out of 12 galaxies, six are star-forming while four galaxies show LINER-like emission. The spectra of the other two galaxies have low S/N and do not show strong emission line features to fit and extract the parameters. We classify them as galaxies with no emission lines. The star-forming nature of six galaxies and the lack of emission in two galaxies support our hypothesis of the evolutionary sequence of bar-driven quenching, with central sub-kpc star-formation as a pre-bar-quenching phase, due to inflow of gas towards the centre, from the action of the bar. Once the fuel for star formation in the nuclear region is exhausted, there is no further fuel for star formation through the bar, and the entire region within the bar co-rotation radius becomes fully quenched. Multiple mechanisms can contribute to the observed LINER-like emission in four of our sample galaxies, including the bar-driven quenching (leading to the presence of old post-AGB stars), bar-driven dynamical processes such as shocks and shear, and LLAGN. The LLAGN may also contribute to internal quenching; the implications of this in our study are discussed in detail in subsection 4.5.
We note that there are other emission line diagnostic methods described in the literature to understand the nature of photoionisation. This includes the use of H$\alpha$ equivalent widths( EW$_{H\alpha}$) to distinguish galaxies into the strong AGN, weak AGN/LINERs, passive, or star-forming systems. We measured the EW$_{H\alpha}$ for our sample and found values ranging from 1.57 - 8.81 \AA\ . The EW$_{H\alpha}$ for most systems is $>$ 3 \AA\, which puts them in either star-forming or AGN (strong/weak) categories, except for galaxies 2 (1.57 \AA\ ) and 4 (2.11 \AA\ ). The low EW$_{H\alpha}$ values in the two galaxies may result from dilution in emission-line measurements due to aperture effects and/or a strong old stellar continuum, which can suppress the observed equivalent widths \citep{2011MNRAS.413.1687C}. \cite{2011MNRAS.413.1687C} also use [NII]/H$\alpha$ ratios (> -0.4) to further classify star-forming (< -0.4) and AGN-like(> -0.4) systems when the EW$_{H\alpha}$ are greater than 3 \AA\ . However, in our sample, most of the systems are classified as composite and have [NII]/H$\alpha$ > -0.4. Previous studies suggest that elevated [NII]/H$\alpha$ ratios may arise from shocks, shear, or galactic winds that can harden the ionising radiation field, and/or from enhanced metallicities often seen in barred galaxies \citep{1994ApJ...430L.105F, 2002ApJS..142...35K, 2011MNRAS.416.2182E, 2014MNRAS.444.3894H,  2015A&A...584A..88F, 2016A&A...595A..63V, Kim_2024}. These effects might be more prominent for our sample, as the fibre spectra probe regions larger than $\sim$ 1 kpc in most cases.
In the next section, we analyse the nature and dominant location of UV emission using spatially resolved UV-optical colour maps, and subsequently determine the age of the stellar populations in different regions of our sample galaxies using NUV-r radial profiles. 

\subsection{Spatially resolved UV-optical colour colour map}
 We construct spatially resolved UV–optical colour maps for all 12 galaxies using SDSS r-band and GALEX FUV and NUV imaging. These maps enable us to distinguish UV emission arising from young stellar populations or from older, hot stellar components \citep{2011ApJS..195...22Y}. To examine spatial variations in the UV–optical colours, we define a set of subregions within each galaxy following the approach described in our previous work in RD25. These regions include a circular bulge region with a fixed diameter of 6", an elliptical bar region whose major axis corresponds to the bar length derived from isophotal analysis of the SDSS r-band image, and other regions such as the bar excluding the bulge, and a circular bar region with a radius equal to half the bar length. The full galaxy is defined using an elliptical aperture with a semi-major axis equal to the optical radius (R$_{25}$) and the disc region as the annulus between the galaxy and circular bar apertures. All regions are overlaid on the degraded SDSS r-band and GALEX FUV and NUV images, matched in spatial resolution. The adopted fixed bulge radius corresponds to physical scales of 1.24-3.39 kpc across our redshift range (Table 1). While this aperture may marginally include portions of the bar or disc in some cases, it remains representative of the inner region typically dominated by bulge light. For comparison, the effective bulge radii reported by \citet{kruk2018galaxy} range from 0.47 to 1.96 kpc, and most systems exhibit S\'ersic indices below 2, consistent with pseudo-bulge morphologies. This suggests that internal quenching in these galaxies is unlikely to be strongly influenced by a classical bulge component. We also take note here that at the available spatial resolution, the bulge region is not fully resolved, and therefore the potential contribution from sub-kpc nuclear/central emission could be diluted and is not directly apparent in the bulge UV–optical colours. \\
\indent As shown in Figure 6, we present two-colour maps for the various subregions defined for our sample galaxies and RD25 sample, allowing us to distinguish between star-forming and passive regions based on their colour properties. The magnitudes in all three bands (FUV, NUV, and r) are measured for each region, corrected for Milky Way extinction, and subsequently used to estimate the colours.
The bulge and bar components predominantly fall within the range NUV–r = 4–5 mag, while the circular bar regions exhibit colours redder than 3.5 mag. In contrast, the annular disc regions show NUV–r colour < 3 mag. This supports the inside-out quenching scenario with bar and bulge regions older and redder in colour compared to the disc region. When compared to the RD25 sample, the colours for inner structures, i.e., the bar and bulge regions, are relatively bluer. This may imply that the corresponding stellar populations are also relatively younger than those in the RD25 sample.

\subsection{NUV-r profiles}
\begin{figure*}[ht!]   
\centering
\includegraphics[width=0.97\linewidth]{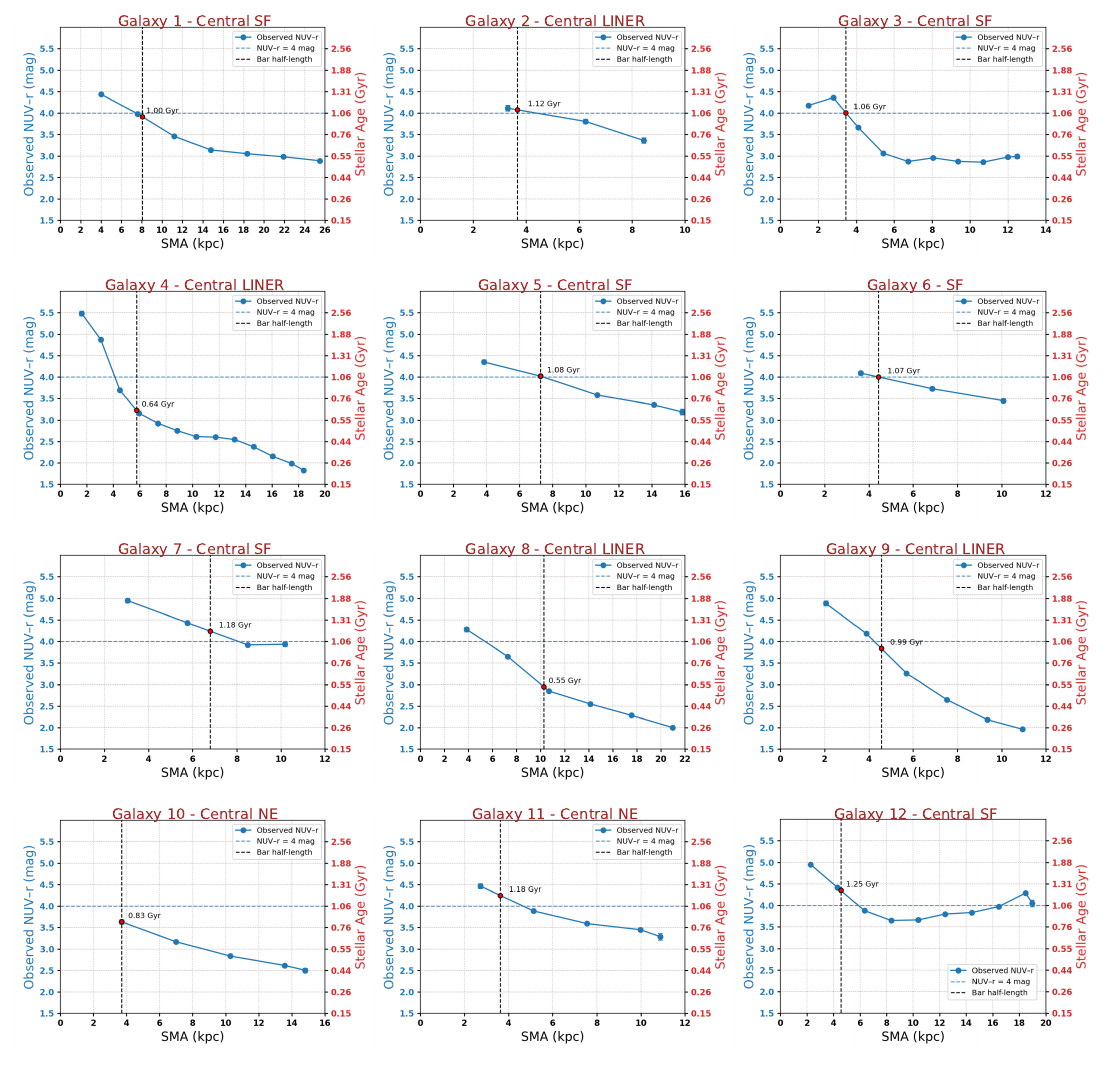}
\caption{NUV-r colour radial profile for each sample galaxy is shown with stellar ages on the other y axis. The blue dashed horizontal line references here NUV-r = 4 mag, while the vertical black line represents the bar half-length. The red dot represents the dominant stellar age at the end of the bar.}
\label{fig7}
\end{figure*}
Figure 7 shows the NUV–r radial color profiles for our complete sample. These profiles are created by placing concentric elliptical annuli, each 3" in width, on the SDSS r-band images (convolved to match the GALEX NUV PSF) and the GALEX NUV images, extending radially out to the optical $R_{25}$ radius. For the six galaxies exhibiting central residual star formation in their SDSS fiber spectra from BPT, all display redder central regions extending up to the length of the bar. All galaxies exhibit bluer outer discs compared to their inner regions, indicating the inside-out quenching. As described in Section 3, NUV-r colour provides a relative measure of stellar population properties. We further translate the observed NUV–r colors into stellar ages by interpolating the theoretical age–color relation from the E-MILES stellar population models. They are indicated on the right y-axis for each galaxy in Figure 7. The corresponding stellar population ages, as shown in Figure 7, are derived from the observed NUV-r color using the synthetic age-color relation (Figure 3). The stellar ages lie in the range of 1.00 – 1.25 Gyr across these six galaxies. All have ages exceeding 1 Gyr, consistent with the observed redder colors within the bar region and indicative of older, more evolved stellar populations despite residual star formation activity in the nuclear region. Here, we have NUV-r profiles produced using GALEX NUV data, which have coarser resolution. However, in higher spatial resolution data, we expect to see a bluer central sub-kpc region and a redder region extending to the ends of the bar. A mild indication of this trend is observed in the profile of Galaxy 3. This galaxy is the nearest in our sample, and the NUV-r profile is sampled at a small physical scale, which may be the reason we observe this effect. If these galaxies are in the evolutionary stage, just before the fully-quenched phase in the bar-driven quenching scenario, we expect the stellar populations within the bar co-rotation region to be slightly younger than the population within the bar co-rotation region of the RD25 sample. This trend will be examined in more detail in a future study of nearby barred galaxies.\\ 
\indent In Figure 8, we present the normalized median NUV–r radial color profiles for our full sample, divided into star-forming and LINER galaxies based on their BPT classifications, and compare them with the sample from RD25. Each data point in the profile represents the median color within a bin size of 0.2 dex, and the shaded regions indicate the interquartile range, bounded by the 25th and 75th percentiles of the color values in each bin. Figure 8a compares the profiles of the six BPT-classified star-forming galaxies with the centrally quenched sample from RD25. The RD25 sample now also includes two additional galaxies - galaxies 10 and 11, which were previously categorized as no emission line galaxies. These six galaxies exhibit slightly bluer NUV–r colours than the RD25 sample, although the values remain above NUV–r = 4 mag, indicating that the stellar populations within the bar region are older than 1 Gyr but younger than those at the respective radii of the RD25 sample. The dashed purple and light brown vertical lines represent the median bar half-lengths for the two samples, and they are found to be similar. In general, a clear transition is observed from redder to bluer colors at the end of the bar region. This implies that the inner region is quenched up to the bar length while discs are still star-forming. Since all these galaxies exhibit star formation in their central fibre spectra region, we infer that these six galaxies may be undergoing their final episode of star formation. \\
\begin{figure*}[h!]   
   \centering
     \begin{subfigure}{0.30\textwidth}
   \includegraphics[width=\linewidth]{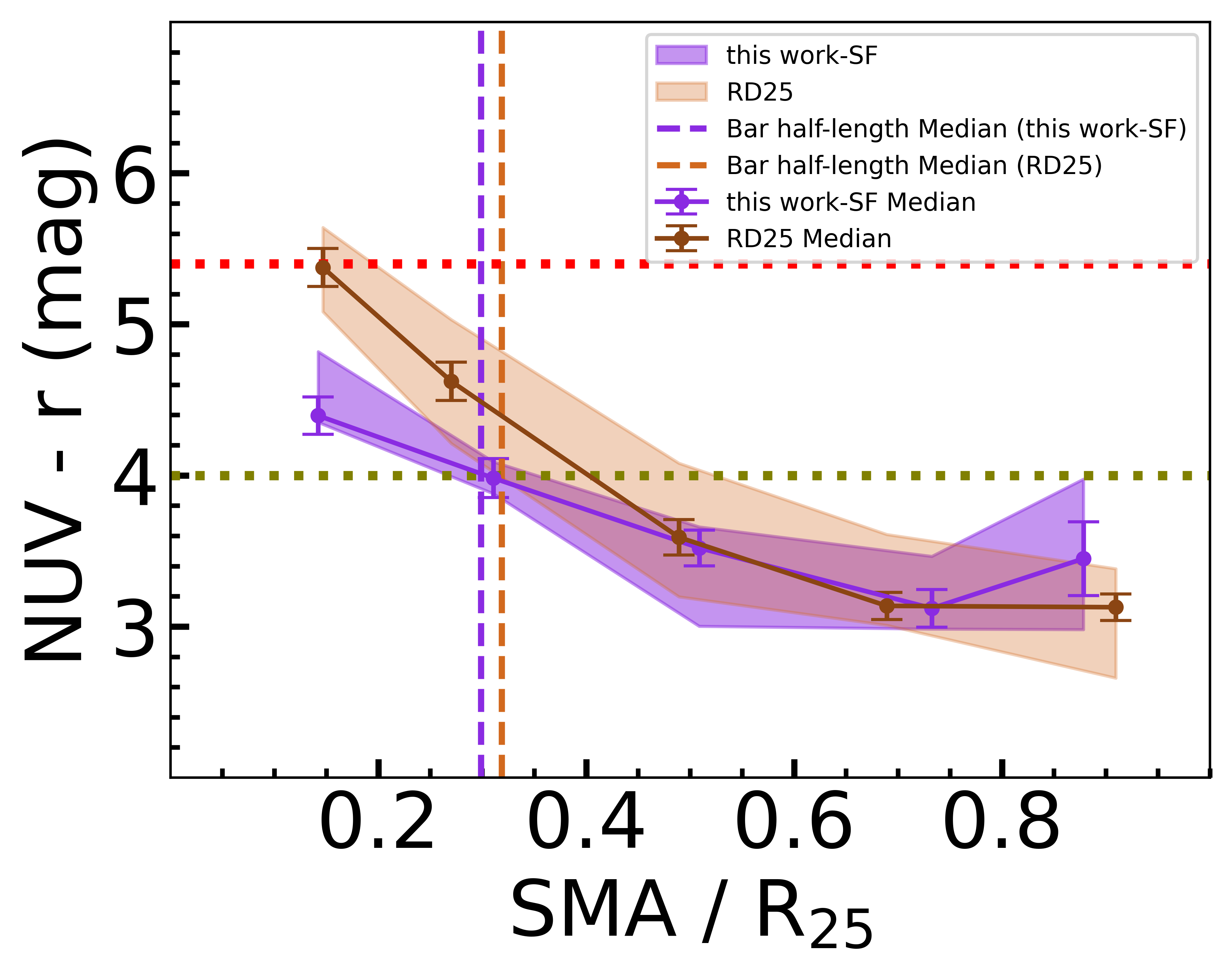}
   \subcaption{}
   \label{fig8a}
\end{subfigure}
\begin{subfigure}{0.30\textwidth}
   \includegraphics[width=\linewidth]{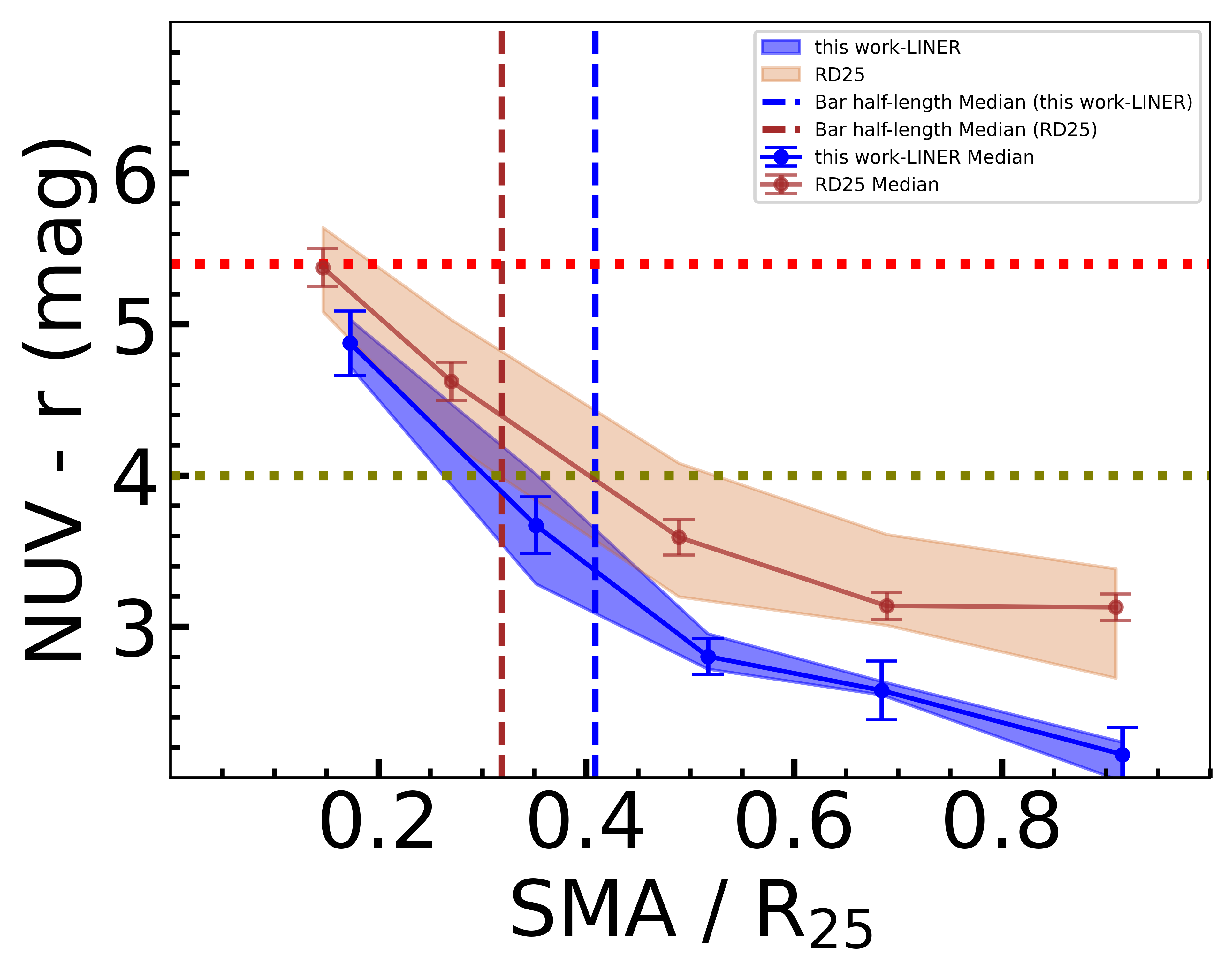}
   \subcaption{}
   \label{fig8b}
\end{subfigure}
\begin{subfigure}{0.30\textwidth}
   \includegraphics[width=\linewidth]{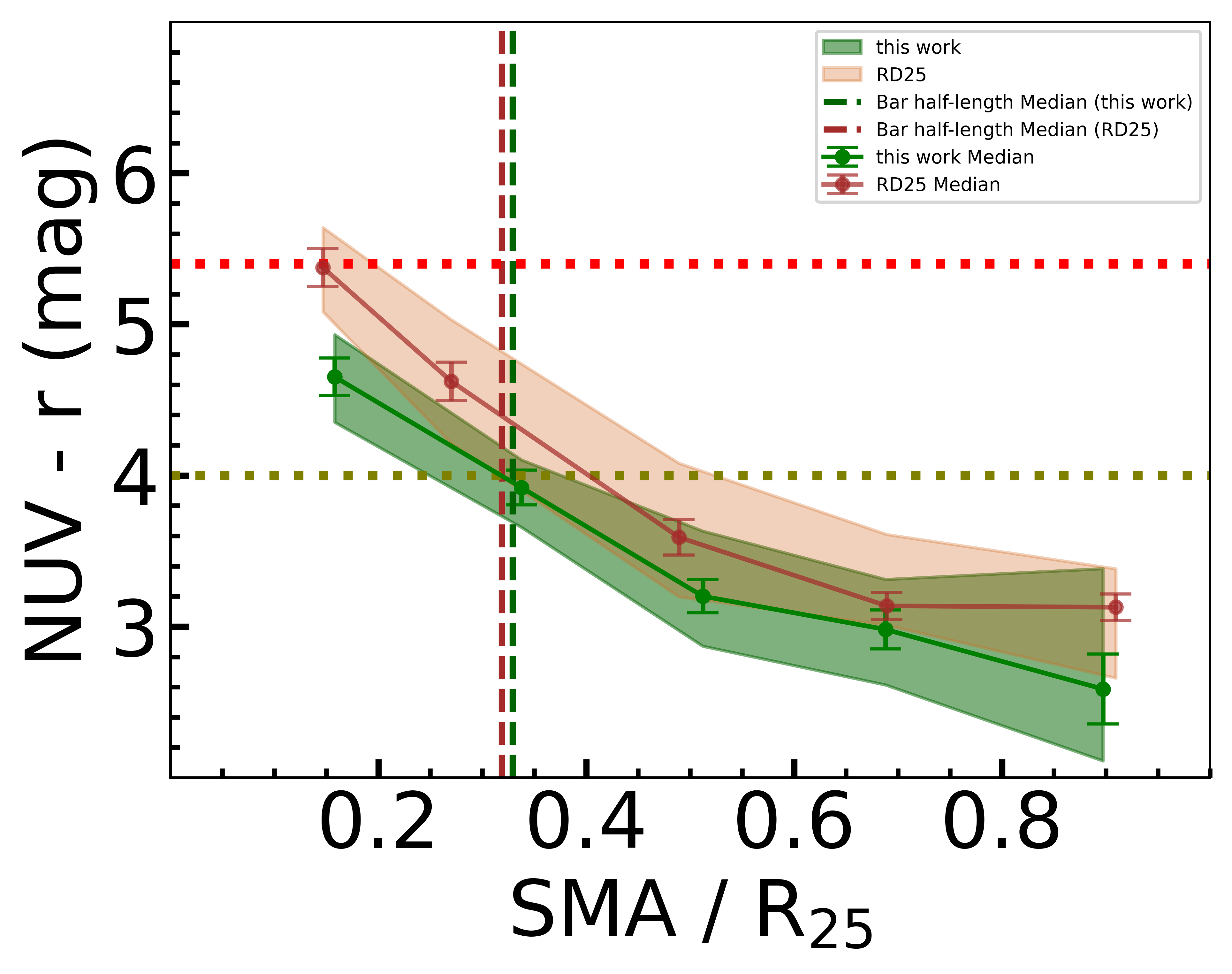}
   \subcaption{}
   \label{fig8c}
\end{subfigure}
\caption{(a) Median NUV–r radial colour profiles for the six galaxies classified as star-forming in BPT-2 (in violet), shown in comparison to the R25 sample (in light brown). The vertical dashed line indicates the bar length for each galaxy. (b) Median NUV–r radial colour profiles for the four galaxies identified as LINERs in BPT-2, also compared with the R25 sample (light brown), with vertical dashed lines denoting the respective bar lengths. (c) Median NUV–r radial colour profiles for all 11 galaxies (green), and compared with the RD25 sample (light brown), with vertical dashed lines denoting the respective bar lengths. Galaxies 10 and 11 are included in the RD25 sample here.}
\label{fig8}
\end{figure*}
\indent The two galaxies, galaxy 10 and 11, which we classify as systems with no detectable emission lines, have their NUV-r colour profiles shown in Figure 7. We consider these galaxies as similar to the RD25 sample. Like RD25 sample, the profile for galaxy 11 is consistent with older stellar population ages of $\sim$ 1 Gyr, with the bar region fully quenched. While for galaxy 10, the colour values in NUV-r profiles are below 4 mag, indicating that the overall bluer color is due to a younger population. We note here that \cite{kruk2018galaxy} has classified this galaxy as a bulgeless galaxy for which only bar and disc components were fitted. In addition, the bar size estimated from ellipse fitting is comparable to the bin size adopted in our analysis and to the region covered, defined as the bulge region. Due to limited resolution, it is therefore difficult to robustly isolate the colour of the bar region and to draw a firm conclusion about the associated stellar population.\\
\indent In Figure 7, galaxies classified as LINERs exhibit a gradual transition from redder to bluer colours, although this transition (NUV–r $<$ 4 mag) does not always coincide with the bar ends. This suggests that the bar regions in these galaxies are not fully quenched to the bar end and host stellar populations younger than 1 Gyr. The dominant stellar population ages for these galaxies range from 0.55 - 1.12 Gyr. Galaxies 4 and 8 show stellar populations ages of around $\sim$ 0.6 Gyr, implying that these galaxies are quenched to a small inner region, well before the ends of the bar. Galaxy 9 hosts stellar populations of around 0.99 Gyr and is quenched right up to the ends of the bar. Galaxy 2 has very few data points for exact interpretation, but it can be seen that inside the bar region, stellar populations are older than 1 Gyr. In Figure 8b, we compare the normalised median profiles of these four LINER-classified galaxies with the RD25 sample. The LINER-classified galaxies display slightly bluer NUV–r colours than the RD25 sample. The median bar length of the LINER-classified galaxies is indicated by the blue vertical line, and the transition from redder to bluer colours seems to occur earlier than the bar length. The bar lengths of LINER galaxies are also larger in comparison to both star-forming BPT classified galaxies and RD25 centrally quenched galaxies. What we understand here is that these four galaxies exhibit varying levels of star formation within their bar regions. This raises the question of which physical mechanisms are responsible for quenching in these LINER-classified galaxies. Recent studies (\citet{2024MNRAS.532.2320G, 2025A&A...699A.204M} and references therein) have shown that bar-driven gas inflows can trigger AGN activity and potentially lead to internal quenching, although the spatial extent of such internal quenching remains poorly constrained. \cite{2025MNRAS.537.3543F}, using TNG50 simulations, demonstrated that barred galaxies can develop 3 -- 15 kpc wide central gas holes from supermassive black hole (SMBH) feedback, but only for systems hosting black holes with masses ($\geq 10^{8} M_{\odot}$). In the next subsection 4.5, we calculate the masses of black holes in our sample and discuss this implication in more detail. \\
\indent Regardless of whether the galaxies are classified as star-forming or LINERs, the bar regions mostly appear redder and quenched, whereas the discs remain bluer and actively star-forming. This is illustrated in 8c, where we compare the normalised median NUV-r color radial profiles for all 12 sample galaxies with the RD25 sample. For the current sample, the NUV–r colours are systematically bluer than the RD25 sample. It is possible that the current sample is approaching the final stage of bar quenching, where the inner regions are fully quenched with only a residual amount of activity remaining, while in the RD25 sample,  the inner regions are fully quenched, having reached the end stage of the bar-driven quenching phase.\\
 We note that all magnitudes have been corrected for foreground Galactic dust extinction \citep{2011ApJ...737..103S, 1989ApJ...345..245C}, but not for internal dust attenuation.
To check for the effect of internal dust extinction, we followed the same steps as in RD25. We used the dust attenuation value A$_{v}$ provided in the GSWLC M2 catalog, which is determined from SED fits constrained with IR luminosity. These values characterize internal extinction and would make the colours slightly bluer (by approximately 0.25 mag). This effect is small and does not alter the overall results. The dust attenuation may also vary with galactocentric radius and, as such, can influence the observed NUV-r colour profiles. Studies such as \cite{2015A&A...581A.103G} and \cite{2016ApJ...817L...9N} have shown that for massive galaxies, the variation in dust attenuation is mainly observed within the central regions ($<$ 1 kpc) and has minimal variation at larger radii. Since the spatial resolution of the NUV-r profiles probed in this study is coarser than 1 kpc, central dust gradients within 1 kpc cannot be resolved and are therefore unlikely to significantly affect the observed profiles. As a result, the effect of internal extinction on the derived colours and the approximate ages of the stellar populations in different regions of our galaxies is expected to be minimal.

\indent We inspect the bar properties, mainly bar length (normalised against disc length, R$_{25}$) and ellipticity, for both our current sample and the RD25 sample. To check whether these properties differ between the two samples, we use the Kolmogorov-Smirnov (KS) test to compare whether they are drawn from the same distribution and the Mann–Whitney U test to evaluate whether the two samples' median values differ significantly. We find no statistically significant difference in the bar lengths (normalised) distribution or in their median values (KS: p $\approx$ 0.43, MW: p $\approx$ 0.28), which indicates that bar lengths are similar in both samples. However, the ellipticity distributions for the two samples show a statistically significant difference for the median values (MW: $\approx$ 0.018), but the overall distributions are only marginally different (KS: p $\approx$ 0.076). This suggests that bar length and ellipticity may not be a dominant parameter controlling the differences between two stages of bar evolution, although a larger sample would be required for more robust inference.

\subsection{Role of AGN feedback}
\label{subsection4.5}
 The low-luminosity AGN activity can lead to the development of central cavities or regions depleted of star formation. Recent studies suggest that the quenching is not primarily driven by the instantaneous impact of AGN activity, but rather by the cumulative power output of integrated AGN feedback. \cite{2022MNRAS.512.1052P} and \cite{2025MNRAS.537.3543F} described a growing black hole mass $\geq 10^{8} M_{\odot}$ that undergoes kinetic feedback and drives cold gas outwards through high-energy winds. These winds, exceeding the binding energy of the host, inject turbulence in the galaxy's ISM and prevent gas from undergoing the gravitational collapse \citep{2020MNRAS.493.1888T, 2020MNRAS.499..768Z, 2021MNRAS.508.4667P, 2023MNRAS.526..217A}. This mode of kinetic feedback has been both predicted in simulations, such as IllustrisTNG, and confirmed observationally in SDSS data. To gauge this operation in our sample, we estimated the black hole masses for all sample galaxies using the black hole mass-sigma ($M_{\rm BH}-\sigma$) relation for late-type galaxies (LTG) from \cite{2013ApJ...764..184M}. 
 $$logM_{BH} = 5.06 \times log(\sigma_{c}/200) + 8.027,$$
 where $\sigma_{c}$ is the central velocity dispersion obtained from pPXF spectra fit and is corrected for aperture and inclination effects \citep{1995MNRAS.276.1341J} using b/a and effective radius ($R_{e}$) values taken from \cite{kruk2018galaxy}.
 
 The black hole masses in our galaxies fall within the range $log (M_{BH}/M_{\odot}) \sim 6.09-7.76$ (also provided in Table 2). We also estimate black hole masses for the RD25 sample,  and they lie in the range $log (M_{BH}/M_{\odot}) \sim 6.46-7.74$ . The uncertainties in the velocity dispersion are obtained from the pPXF analysis, while the black hole mass uncertainties are derived through error propagation. These estimated masses lie below the critical threshold of $log (M_{BH}/M_{\odot}) \sim 8.0$, above which AGN feedback is expected to dominate over bar-driven quenching. This result suggests that AGN activity is unlikely to be the primary driver of quenching in our current and RD25 sample, although it could explain the LINER-like ionisation signatures observed. These estimated black hole masses also rule out the possibility of any past AGN activity in our sample galaxies. This instead points to ionisation by hot evolved stars or by shocks and shear, mechanisms that are often linked with bar-driven processes.

\section{Discussion}
In this study, we investigated a sample of 12 centrally quenched barred galaxies that exhibit central emission in their SDSS fibre spectra. For half of the sample, BPT diagnostics revealed that this central emission originates from ongoing star formation. The NUV-r colour profiles of these galaxies showed that the region inside their bar co-rotation radius (region up to the bar ends) is redder (NUV–r > 4 mag) in colour, indicating an older stellar population with ages older than 1 Gyr. These findings are consistent with the results of \citet{2009A&A...501..207J} and \citet{2015MNRAS.450.3503J}, where they proposed the formation of a 'star formation desert' within the bar-covered region, except for the innermost nuclear zone. \citet{2019MNRAS.489.4992D} studied six simulated barred galaxies and showed that, in the absence of contamination from radial stellar migration, the star formation desert hosts stellar populations with ages truncated at approximately 1 Gyr. Bars play a crucial role in redistributing gas and stars within the disc. When gas comes into contact with the leading edge of the bar, it experiences strong shocks and is transported inward along the dust lanes, often seen tracing the bar. This inflowing gas accumulates in the nuclear regions, triggering intense nuclear starbursts. \cite{2019MNRAS.489.4992D} suggests that this process unfolds within $\sim$1 Gyr of bar formation \citep{2019MNRAS.489.4992D}, depleting the region encompassed by the bar of fuel for star formation. At the same time, the bar-driven non-circular motions increase the velocity dispersion of the gas clouds, making the gas more turbulent and stabilizing it against gravitational collapse, thus suppressing further star formation effectively in a period of 1 Gyr \citep{khoperskov2018bar}. Either mechanism or the combination of both can effectively lead to a morphological change in which a disc galaxy becomes centrally quenched, with star formation restricted to sub-kpc nuclear regions and at the bar ends. \citet{2015MNRAS.446.2468E} and \citet{2015MNRAS.454.3299R} also demonstrated the suppression of star formation in the bar interior, driven by the high shocks and shear experienced by the gas. The central region may exhibit episodic nuclear starbursts and eventually form nuclear rings. Similar conclusions were reached by \citet{spinoso2016bar}, although they suggested that bar-driven quenching results from the inward transport and subsequent depletion of gas. Through an analysis of recent star formation and gas redistribution of four barred galaxies, \citet{george2020more} showed that bars can effectively funnel gas inward. In three of these galaxies, the bar region was entirely devoid of star-forming gas, with molecular gas confined to the central sub-kpc region. The fourth galaxy appeared to be in a transitional phase, still exhibiting recent star formation along the bar as gas redistribution is still in process. Studies such as \citet{2007A&A...474...43V}, \citet{2020MNRAS.495.4158F} and \cite{2020A&A...644A..38D} analyzed the H$\alpha$ morphology of barred galaxies and found varying levels of star formation activity within the bar region and near it. \\
\indent We find that for both our current sample and the RD25 sample, the stellar populations within the bar region are older than 1 Gyr. However, the colours of our current sample are bluer, implying younger stellar populations than in RD25. This difference suggests that the two samples may represent distinct evolutionary stages of bar-driven quenching. In particular, ongoing star formation is clearly detected for the current sample in the SDSS fibre spectra but not for RD25, indicating that the current sample is likely in a pre-quenching phase that precedes the evolutionary state of the RD25 systems. Based on our results and other findings in the literature, a scenario is possible in which, shortly after bar formation, gas is funneled inward, leading to star formation along the bar, in its surroundings, as well as in the center. Over Gyr timescales, star formation within the bar region declines as the gas is depleted or stabilised, while the central gas undergoes a final starburst that eventually fades \citep{james2018star, 2019MNRAS.489.4992D,spinoso2016bar, khoperskov2018bar, george2020more, 2024A&A...687A.255S}. The result is a quenched inner disc up to the bar end, with star formation confined only to the outer disc.\\
\indent The observed LINER-like emission in our sample may represent an intermediate or later stage in this bar-driven evolutionary sequence. As discussed in subsection 4.2, several mechanisms, such as low-luminosity AGN activity, ionization by hot evolved stars, or shock excitation, can contribute to the observed LINER signatures. However, since the estimated black hole masses for these galaxies lie below the critical threshold required to sustain significant AGN activity, these LINER features are more likely linked to ionization from hot evolved stars or shocks. Our analysis suggests that the LINER-classified galaxies in our sample are likely associated either directly or indirectly with bar-driven quenching processes, potentially representing the final phases of bar-induced evolutionary quenching.\\
\indent It is also necessary to point out that, beyond bar-driven secular evolution, several alternative mechanisms can also contribute to nuclear star formation in galaxies, even in those hosting bars. In particular, gas-rich mergers and tidal interactions can induce nuclear starbursts by driving gas inflows through strong gravitational torques. Such processes often lead to enhanced central SFRs \citep{1996ApJ...464..641M, 2011MNRAS.416.2182E, 2011A&A...533A.104E, 2015A&A...573A..78Q, 2015MNRAS.452.3551W, 2018MNRAS.473.2521S, 2020A&A...635A.197D, 2022ApJ...941..128P}. These interactions can also sometimes trigger instabilities and aid the formation of bars that further aid in the gas inflows and nuclear star formation \citep{2021A&A...650A.191S, 2025A&ARv..33....7S}. \cite{2008A&ARv..15..189S} discusses how cold gas accretion can sustain star formation, where gas infall onto central regions can directly happen through either minor mergers or galactic fountains maintained through the feedback cycle. However, most of these processes leave observable signatures such as tidal tails, loops, warped disks, or other disturbed morphological features. But for nuclear processes governed through only secular evolution, bar should be the primary driver \citep{2011MNRAS.416.2182E, 2020A&A...643A..14G}. A visual inspection of our sample galaxies does not reveal clear signatures of disturbed morphology or extended tidal structures. However, this is based on SDSS imaging alone, and only deeper observations may be able to fully rule out faint interaction features.

\indent Thus, for our sample, whether the nuclear regions exhibit ongoing star formation or LINER-like emission, both likely trace different evolutionary stages of bar-driven quenching. Since our study is limited by the coarse spatial resolution of GALEX, this restricts our ability to probe the inner structures of barred galaxies in detail. Higher-resolution data are essential for resolving the central regions more effectively. Integral Field Unit (IFU) spectroscopy represents the current state-of-the-art approach for such investigations. In future work, we aim to utilize IFU data to examine the inner regions of barred galaxies with higher spatial resolution. Additionally, we plan to extend our analysis to galaxies occupying different regions of the star formation rate–stellar mass (SFR–M$_{*}$) plane, in order to gain a better understanding of the evolutionary stages of bar-driven quenching.

\section{Summary}
In this study, we examined centrally quenched barred galaxies in the redshift range $0.01 < z < 0.06$ that showed residual central emission within the SDSS fiber spectra. Our main conclusions are as follows:
\begin{enumerate}
\item We identify 12 barred galaxies as centrally quenched, those with extended star-forming discs but quenched inner regions, by leveraging the differences in star formation rates between the MPA-JHU and GSWLC catalogs. However, these galaxies also host residual central emission within the SDSS fiber spectra.
\item All but one galaxy in the sample hosts pseudo-bulges, suggesting that the central regions have predominantly evolved through secular processes.
\item Emission-line fitting of the fibre spectra using the pPXF code shows that the central emission arises from either ongoing star formation or LINER-like ionization, suggesting diverse ionisation mechanisms.
\item  Spatially resolved UV-optical two colour diagrams reveal that the bulge and bar regions are dominated by redder colours (NUV - r > 4 mag), consistent with older stellar populations, while the surrounding discs are bluer (NUV - r < 4 mag), indicating younger populations.
\item Median NUV–r radial colour profiles show a systematic transition from redder to bluer colours around the bar length. In star-forming systems, this transition coincides with the end of the bar with dominant median stellar age inside the bar region $\sim$ 1.08 Gyr, whereas in LINER-classified galaxies, it occurs within the bar length with corresponding stellar ages of $\sim$ 0.82 Gyr.
\item The median NUV-r colour profile of our current sample is bluer with a corresponding younger population when compared to the RD25 sample studied in our previous work at the same radii, suggesting they are different evolutionary stages in bar-driven quenching.
\item The relatively low black hole masses in our sample ($log (M_{BH}/M_{\odot}) < 8.0$) suggest that kinetic-mode AGN feedback is unlikely to play a dominant role, making an AGN-driven origin of the LINER activity or a past AGN less probable. Instead, this supports a picture in which bar-driven processes play the primary role in regulating quenching.
\item We interpret that the star-forming classified galaxies are in the final stages of bar-driven quenching (pre-quenching phase), where residual star formation is confined to the central kiloparsec and the bar region is largely quenched. The LINER systems may also represent quenching end states, depending on the physical origin of their LINER-like emission, which is mostly also an indirect outcome of bar-driven processes.
\item To resolve the sub-kpc central region, higher spatial resolution imaging and spectroscopic observations are required, which are essential for constraining both the evolutionary sequence of bar quenching and the origin of LINER emission.
\end{enumerate}

\begin{acknowledgements}
SS acknowledges support from the Alexander von Humboldt Foundation. This study utilized archival Sloan Digital Sky Survey (SDSS) Data Release 7 (DR7) and Galaxy Evolution Explorer (GALEX) Medium Deep Catalogues data. Funding for the SDSS and SDSS-II was provided by the Alfred P. Sloan Foundation, the Participating Institutions, the National Science Foundation, the U.S. Department of Energy, the National Aeronautics and Space Administration, the Japanese Monbukagakusho, the Max Planck Society, and the Higher Education Funding Council for England. The SDSS was managed by the Astrophysical Research Consortium for the Participating Institutions. GALEX (Galaxy Evolution Explorer) is a NASA Small Explorer, launched in 2003 April. We gratefully acknowledge NASA’s support for construction, operation, and science analysis for the GALEX mission, developed in cooperation with the Centre National d’Etudes Spatiales of France and the Korean Ministry of Science and Technology. This study also made use of Astropy, a community-developed core Python package for Astronomy (Astropy Collaboration, 2013, 2018). 
\end{acknowledgements}
\bibliography{reference}

\end{document}